\documentclass[twocolumn,aps,prb,superscriptaddress]{revtex4-1}

\usepackage[english]{babel}

\usepackage{graphicx}
\usepackage{amsmath}
\usepackage{amsfonts}
\usepackage{amssymb}
\usepackage{bm,bbm}
\usepackage{color}
\usepackage{hyperref}
\usepackage[shortlabels]{enumitem}

\newcommand{\sx}[1]{\sigma^{\rm x}_{#1}}
\newcommand{\sy}[1]{\sigma^{\rm y}_{#1}}
\newcommand{\sz}[1]{\sigma^{\rm z}_{#1}}
\newcommand{\p}[1]{\sigma^{+}_{#1}}
\newcommand{\m}[1]{\sigma^{-}_{#1}}
\newcommand{\ssx}[1]{s^{\rm x}_{#1}}
\newcommand{\ssy}[1]{s^{\rm y}_{#1}}
\newcommand{\ssz}[1]{s^{\rm z}_{#1}}
\def\ii{{\rm i}}
\newcommand{\ket}[1]{|{#1}\rangle}

\newcommand{\bracket}[3]{\langle{#1}|{#2}|{#3}\rangle}

\newcommand{\ave}[1]{\langle{#1}\rangle}

\def\tit#1{{\em #1},}
\def\etal#1{#1}

\def\Dn{D_{\rm NESS}}
\def\Du{D_{\rm unit}}

\begin{document}

\title{Diffusive transport in a quasiperiodic Fibonacci chain: absence of many-body localization at small interactions}

\author{Vipin Kerala Varma} 
\affiliation{Department of Physics and Astronomy, College of Staten Island, CUNY, Staten Island, NY 10314, USA}
\affiliation{Physics program and Initiative for the Theoretical Sciences, The Graduate Center, CUNY, New York, NY 10016, USA }
\affiliation{Department of Physics and Astronomy, University of Pittsburgh, Pittsburgh, PA 15260, USA}

\author{Marko \v Znidari\v c}
\affiliation{Physics Department, Faculty of Mathematics and Physics, University of Ljubljana, Jadranska 19, SI-1000 Ljubljana, Slovenia}

\begin{abstract}
 We study high-temperature magnetization transport in a many-body spin-1/2 chain with on-site quasiperiodic potential governed by the Fibonacci rule. In the absence of interactions it is known that the system is critical with the transport described by a continuously varying dynamical exponent (from ballistic to localized) as a function of the on-site potential strength.
 Upon introducing weak interactions, we find that an anomalous noninteracting dynamical exponent becomes \textit{diffusive} for \textit{any} potential strength. This is borne out by a boundary-driven Lindblad dynamics as well as unitary dynamics, 
 with agreeing diffusion constants. This must be contrasted to random potential where transport is subdiffusive at such small interactions. Mean-field treatment of the dynamics for small $U$ always slows down the non-interacting dynamics to subdiffusion, and is therefore unable to describe diffusion in an interacting quasiperiodic system.
 Finally, briefly exploring larger interactions we find a regime of \textit{interaction-induced} subdiffusive dynamics, despite the on-site potential itself having no ``rare-regions''.
 
\end{abstract}

\date{\today}

\maketitle

\section{Introduction}
How are transport and localization properties altered when correlations are introduced to local fields, both in free fermionic and, particularly, in interacting systems? 
This is the question we address in this paper, focussing on a class of quasiperiodic systems. 
Indeed, in experimental cold atom systems, quasiperiodic systems are more easily realized and have been routinely investigated to address both these aspects \cite{Roati08, Modugno11, exper, Luschen17, Luschen17b}
\par
A textbook one-dimensional (1D) system that is able to describe many gross features of real materials is the tight-binding model~\cite{merminbook}, 
where particles (e.g., electrons) can hop only between nearest neighbor lattice sites. 
An on-site potential can drastically influence the nature of eigenstates, and as a consequence the dynamics of the system: 
an uncorrelated potential can turn extended states in the clean system to localized states \cite{Anderson}, whereas correlations in the potential can either turn the states to critical 
(like in the Fibonacci model) \cite{kohmoto83,ostlund83,kitaev86,sutherland87} or induce a field-strength induced transition between extended and localized states (Aubry-Andre-Harper, or AAH, model) \cite{Harper, AubryAndre}.
An important question is what happens to these states and phases, be it in a random or a quasiperiodic potential, in the presence of interactions. 
In particular, can localization survive~\cite{Fleishman80}, and, if so, under which conditions? And if not, how do the erstwhile extended states now transport conserved quantities?

The persistence of localization in the presence of interactions $-$ many-body localization (MBL) $-$ is considered well established in a range of systems \cite{rew,rew3,rew4}.
There is also mounting evidence that, in a paradigmatic XXZ model with random fields~\cite{foot3}, prior to the MBL transition 
slow subdiffusive magnetization dynamics sets in and continuously slows down to a complete stoppage of transport as the transition point is crossed~\cite{agarwal15, reichman15, sarang15, lea15, PRL16, auerbach16, luitz16, jacek17, doggen18}.
Note that extracting asymptotics of transport in the ergodic phase can be difficult ~\cite{evers17,osor16,robin18}, see also our comments later on in the text. 
What is less established are the critical properties of the transition itself, like its universality class~\cite{khemani17}. For instance, presently employed renormalization group schemes 
\cite{altman15, sid15, romain19} of merging fully ergodic and fully localized blocks might predict different critical properties of the transition if the blocks are allowed to be subthermal; 
the presence of such blocks can potentially change or slow down the ``avalanche'' instabilities of localized regions against non-localized bubbles~\cite{belgians1,belgians2,Roeck17}.
This is because in a subthermal block one has a scaling relation $x \sim t^\beta$ with $\beta<\frac{1}{2}$ between length and time, 
 meaning that the thermalization time scales at least as $\sim L^{1/\beta}$, i.e., it is very long for small $\beta$.

 In particular, by considering quasiperiodic systems we may test the heuristic explanation for subdiffusion as arising from ``rare-regions''~\cite{agarwal15,sarang16,Adam18} 
 -- regions of abnormally high potential gradients -- that in 1D systems can function as bottlenecks to transport, resulting in subdiffusion even in the thermal phase. 
In the absence of rare regions (like in quasiperiodic systems), however, one therefore expects only diffusive transport all the way to the MBL transition.
 Indeed for the interacting AAH model ~\cite{Michal14,Pilati15,Vieri15,Pilati17,Vieri17,Khemani17,Bera17,Lee17,Pixley17,Naldesi,pnas18,kuba18,doggen19,xu19}, this was to a certain extend observed \cite{Pixley17, pnas18}, see however e.g. Refs.~\onlinecite{xu19,Luschen17b}.
In particular, the analysis of transport for small interactions $U$ reveals~\cite{pnas18} that the transition at $U=0$ is discontinuous, and that there is no localized phase around a non-interacting transition 
at $h_c=2$ (as opposed to previous continuity-based suggested phase diagrams at small-$U$ ~\cite{Vadim13,exper}).
Such qualitatively different nature of the ergodic phase might also affect the universality class of the MBL transition in 
 quasiperiodic systems; indeed the presence of multiple MBL universality classes has been suggested from numerical evidence \cite{Khemani17, Zhang18}. There is also evidence that the structure of l-bits deep in the MBL phase are fundamentally changed due to certain types of quasiperiodicity \cite{Alet18}; this implies that procedures to 
extract them \cite{rademaker2016, pekker2017, varma2019} will likely also need to be modified to reliably extract the new structure of l-bits.

Motivated by the above considerations we address the question of transport in the Fibonacci quasiperiodic model. 
There are several reasons rendering this an interesting undertaking. First, there is a general question of how correlations, e.g. random vs. quasiperiodic potential, influence transport and MBL. 
However, also within quasiperiodic systems the Fibonacci model offers several important advantages going beyond the much more commonly studied AAH model. 
The noninteracting model is critical, showing eigensystem (multi)fractality \cite{kohmoto83,ostlund83,kitaev86,sutherland87,jagannathan16}, 
for any potential amplitude and therefore serves as one of the simplest deterministic systems with anomalous transport~\cite{hiramotoabe} in closed and open setting~\cite{VarmaZnidaricPRE}. We note that (multi) fractality is typical at Anderson transitions~\cite{mirlinrmp}. The natural question that arises is: how much do these anomalous transport properties that arise from eigenspectrum fractality (and not a priori rare-regions) persist upon introducing weak interactions to the system? At low temperatures the interacting model has been studied using bosonization, finding~\cite{vidal99,vidal01} an anomalous transport with the scaling exponent depending on the interaction strength and the position of Fermi level. On the other hand at high temperatures, studed in the present paper, and at strong interaction Ref.~\onlinecite{Alet18} interestingly found signs of nondiffusive transport and an MBL transition despite the noninteracting problem having no localized phase. 
\begin{figure}[tp!]
\centerline{\includegraphics[width=0.7\linewidth]{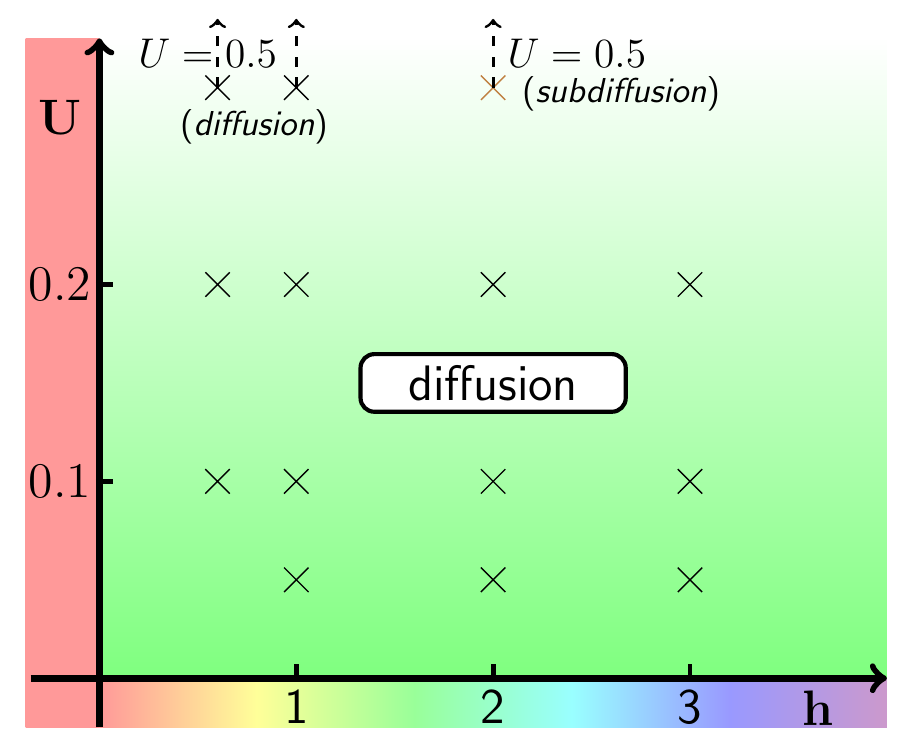}}
\caption{Phase diagram of the Fibonacci model for small $U$: for $h=0$ one has ballistic transport (vertical red bar along the $y$ axis), in the noninteracting limit $U=0$ one goes continuously from ballistic for $h=0$ to localized at $h \to \infty$ (horizontal rainbow bar along the $x$ axis). For nonzero (small) $U$ one has diffusion (green), except possibly at larger $U$ (we have one subdiffusive data point at $U=0.5$ and $h=2$). Crosses mark parameter values for which we gathered data, see Table~\ref{tab1}.}
\label{fig:phase}
\end{figure}

In the present paper we study transport in the interacting Fibonacci model for small interactions at high temperature and half-filling (zero magnetization). 
The phase diagram obtained is shown in Fig.~\ref{fig:phase} . For small interactions $U$ and available potential strengths $h \le 3$ we always find diffusion. 
Anomalous transport discontinuously breaks down to diffusion for any nonzero $U$. 
This is similar as in the AAH model~\cite{pnas18}, with some differences though. One is for instance that at larger potential $h$ transport is qualitatively faster in the Fibonacci model than in the AAH model.  
At larger $U=0.5$ and $h=2$ we, perhaps surprisingly, also observe subdiffusion. This is at variance with a rare-region explanation of subdiffusion but roughly in-line with Ref.~\onlinecite{Alet18} where certain similarity is 
observed between the AAH and Fibonacci models at large $U$. Considering a common discrepancy between different transport studies, a special care is taken using large systems $L \lesssim 1000$ and times $t \lesssim 1000$ to show convergence of two different methods to the same transport.
We also show that a mean-field treatment, that would argue for a dephasing-like explanation of diffusion, is in fact subdiffusive and therefore fails to correctly describe dynamics in the Fibonacci model at small $U$.

\section{Noninteracting Fibonacci model}

 The noninteracting Fibonacci model is described by the Hamiltonian:
 \begin{equation}
H=\sum_{k=1}^{L-1}(\ssx{k}\ssx{k+1}+\ssy{k}\ssy{k+1})+\sum_{k=1}^L \frac{h_k}{2} \ssz{k},
\end{equation}
with a Fibonacci sequence potential, $h_k =2h V(kg)-h$, where $g=(\sqrt{5}-1)/2$ and periodic $V(x):=[x+g]-[x]$ with $[x]$ being an integer part of $x$. Beginning of the sequence is $h_k=h(+1,-1,+1,+1,-1,+1,-1,+1,\ldots)$.

To study transport we initialize the system with a delta function at the central site and compute the mean-squared displacement 
\begin{equation}
 \Delta x^2 (t) = \sum_x \left[x - (L+1)/2\right]^2 |\langle x | \psi(t) \rangle|^2, 
\end{equation}
where 
\begin{equation}
 |\psi(t) \rangle = \textrm{exp}(-\textrm{i}t H) |\psi(0)\rangle,
\end{equation}
is the unitarily evolved initial state with the Fibonacci Hamiltonian. 
We employ a dynamical fit $\Delta x^2 \sim t^{2\beta}$ in order to discern the rate of transport: $\beta = 1$ implies ballistic transport, $\beta = 0$ signifies no transport, 
and $\beta = 1/2$ denotes diffusive transport. 
This is in line with the approach of earlier works \cite{hiramotoabe, Piechon, VarmaZnidaricPRE} but is undertaken here more systematically (larger system sizes, 
as well as inclusion of comparison of finite-size effects).
Secondly, we also consider a new set of initial conditions such that the system is far from linear response. In particular, the system is initialized with a 
fully polarized domain wall (DW) pure state
\begin{equation}
 |\psi(0)\rangle =  \ket{{\rm DW}}=\ket{\uparrow \cdots \uparrow \downarrow \cdots \downarrow},
\label{eq:DWnonint}
\end{equation}
where up/down arrows indicates spins initialized in up or down direction in $z$-direction. 
Here the dynamical rate is quantified by measuring the change in total spin on either half of the chain:
\begin{equation}
 \Delta Z \sim t^{\beta_{DW}}.
\end{equation}
We will find that $\beta_{DW} = \beta$, therefore we will use only one symbol from hereon.
Note that differently from Ref. \onlinecite{hiramotoabe} we do not consider an inversion symmetric potential as a special, and only, sample: 
we have checked that for dynamics on this sample the exponent only slightly increases by about 5\% compared to the values we have presented.
Indeed the inversion symmetric potential for the AAH model too is special \cite{VarmaZnidaricPRE} except that there it instead gives a slightly \textit{smaller} dynamical exponent than for the case of 
averaged data or a generic sample.

Let us first summarize the known physics of the model.
In the Fibonacci model the potential is binary, with only two values $\pm h$, with a quasiperiodic spatial structure following a Fibonacci sequence, $h_k=h(2 V(k g)-1)$, where $g=(\sqrt{5}-1)/2$ and 
periodic $V(x):=[x+g]-[x]$ with $[x]$ being an integer part of $x$.
Compared to the AAH model the potential has a broad momentum spectrum~\cite{monthus17} (instead of a single Fourier component) and is therefore at the other extreme end~\cite{Alet18} of the 
Fourier spectrum broadness (it has as slowly decaying spectrum as possible, compared to as sharply localized one as possible for the AAH model). 
One of the consequences is that the noninteracting version is critical~\cite{kohmoto83,ostlund83,kitaev86,sutherland87} for any $h$. 
The noninteracting Fibonacci model displays a full spectrum of transport scaling exponents~\cite{hiramotoabe, VarmaZnidaricPRE}, going from ballistic $\beta=1$ for $h=0$ to localized $\beta \to 0$ for $h \to \infty$. 
By taking the interacting version we can therefore study, besides the above mentioned effects that a quasiperiodic potential has, also the question of stability against interactions of an arbitrary 
anomalous transport. \par
Let us now explain the technicalities of the model. The Fibonacci sequence required for the Fibonacci model may also be constructed from two symbols ${F,S}$ by the substitution rule
$
 \left(
 \begin{array}{c}
 F \\ S
 \end{array}
 \right)
 \rightarrow
 \left(
 \begin{array}{cc}
 1 & 1 \\ 1 & 0
 \end{array} \right)
  \left( \begin{array}{c} F \\ S \end{array} \right).
$
The transformation matrix has the eigenvalues $g, 1/g$.
%
Repeated application of the above rule 
gives the series of Fibonacci sequences: 
\begin{equation}
\label{eq: FibSeq}
 \{F, FS, FSF, FSFFS, FSFFSFSF, \ldots \}.
\end{equation} 
\begin{figure}
 \includegraphics[width=1.\linewidth]{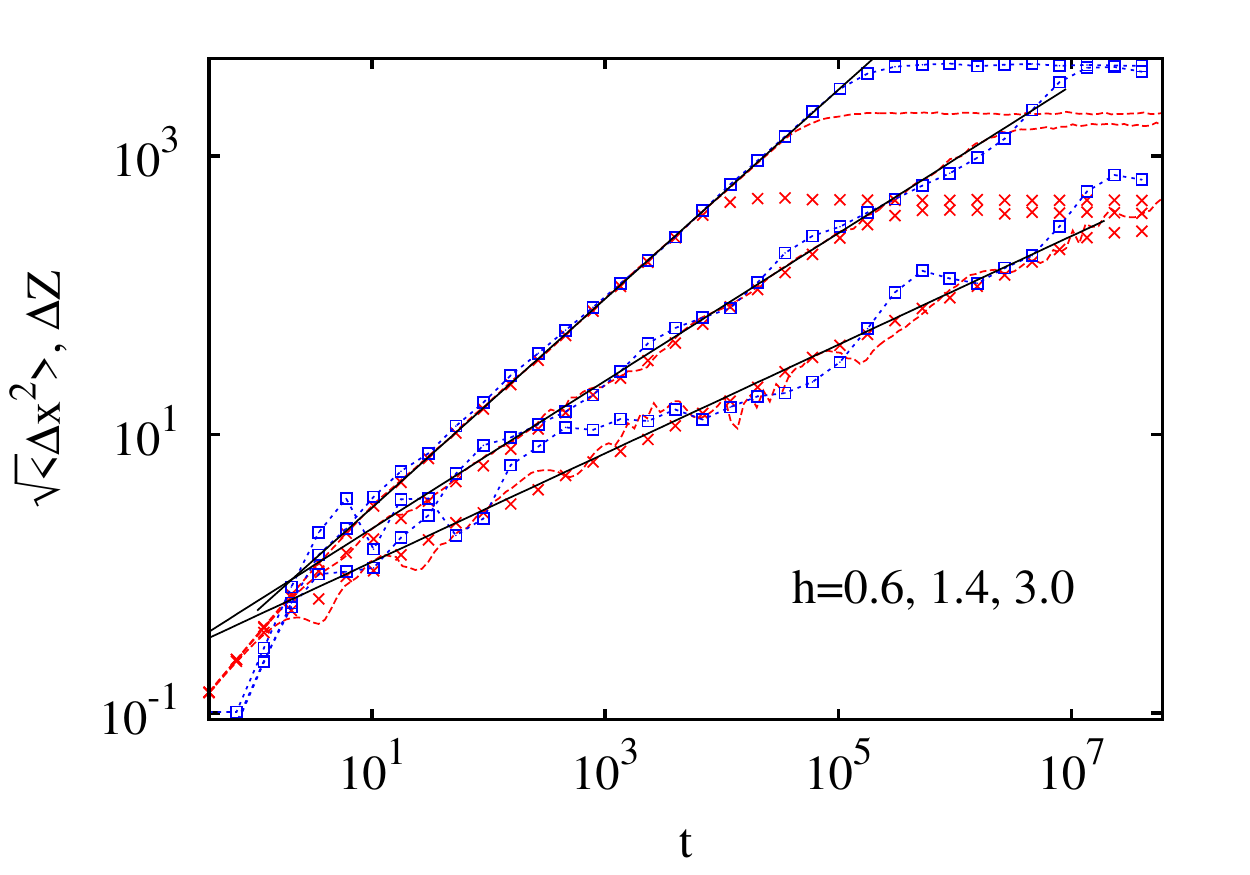}
\includegraphics[width=1.\linewidth]{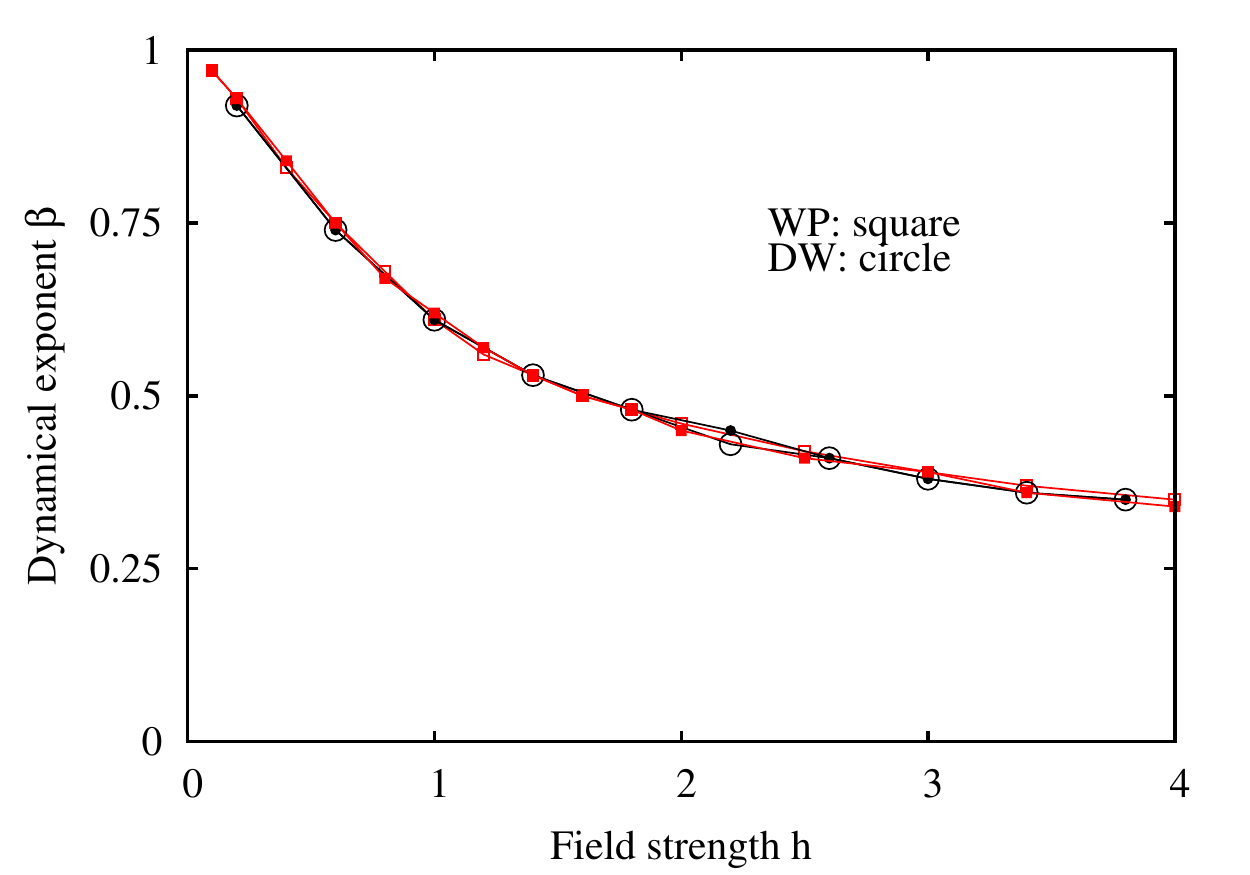}
 \caption{Top: Dynamics of spreading of localized wavepacket (dashed-lines) and half-chain domain wall (blue squares) on $L=8001, 8000$ chain, respectively, for various field strengths $h$, 
 increasing from top to bottom for a single sample;
 the crosses indicate data for $L=2001$ system averaged over 50 samples.
 As $h$ increases, the dynamics gets slower and is anomalous for any nonzero $h$. Moreover the log-periodic oscillations also increases as $h$ is increased. Only for infinite $h$ is the chain localized. 
 Note that the domain-wall data have been scaled by arbitrary constants $\mathcal{O}(1)$ in order to highlight the equality of dynamical exponents more clearly.
 Bottom: Dynamical exponent extracted from all such data in top panel, both for localized wavepacket (WP) and half-chain domain-wall (DW) initial conditions; 
 open symbols are from $L=8000, 8001$ single-sample data whereas closed symbols are from $L=2000, 2001$ averaged data.
 it shows a smooth analytic variation with $h$, interpolating between ballistic $\beta = 1$ and localization $\beta = 0$.
 }
 \label{fig: free}
\end{figure}
Note that the length of each sequence is a Fibonacci number: $1,2,3,5,8 \ldots$, and that, by construction, any sequence always starts with the sequence of any smaller one.
A given cut of length $L$ of any (sufficiently long) Fibonacci sequence determines one sample or configuration of the quasiperiodic chain of length $L$, where 
the quasidisorder potential $h_k$ on site $k$ takes the value $\pm h$ depending on whether the symbol on that site is 
$F$ or $S$ respectively; note that the full long sequence is of Fibonacci length but that need not be true for $L$, the system size under study. 
Moreover, unlike disordered systems where infinite samples exist even for finite systems, here only finite number of samples exist: for instance, on a 3-site chain, the only allowed configurations are 
$FSF, FFS, SFF, SFS$ i.e. $L+1$ configurations in general. However, note that the second and third configurations are reflections of each other, reducing the effective number of independent configurations to half of that. \par
In Fig. \ref{fig: free} we display the spread of an initial wavepacket that is localized at the chain centre and the transfer of magnetization across the centre of the chain when it is initialized to a domain wall 
state (left half of chain has magnetization -1/2, right half has magnetization +1/2 at time $t=0$). We find that the dynamics is identical (top panel), implying equality of dynamical exponents (bottom panel) for 
these two initial states. 
As has been found previously for localized wavepacket spreading \cite{hiramotoabe, Piechon, VarmaZnidaricPRE} there is a continuous decay of the exponent $\beta$ with the field strength. 
Additionally, we find here that the transport rate is robust and remains the same even for bulk excitations such as a domain wall inhomogeneity in the system. For $h=0.6, 1.0, 2.0$, and $3.0$, that we study in the remainder of the paper, we find $\beta=0.75, 0.61, 0.46, 0.39$, respectively. 
Diffusive $\beta=1/2$ is in particular achieved at $h_{\rm diff} \approx 1.6$.

In the Appendix A we in addition show convergence with system size, self-similar structure of a non-interacting $\psi(t)$, and the effects of averaging over different potential instances.

\section{Interacting Fibonacci model}

The interacting version of the Fibonacci model is given by including nearest-neighbour spin-spin interaction terms:
\begin{equation}
H=\sum_{k=1}^{L-1}(\ssx{k}\ssx{k+1}+\ssy{k}\ssy{k+1}+U\ssz{k}\ssz{k+1})+\sum_{k=1}^L \frac{h_k}{2} \ssz{k},
\end{equation}
with the same Fibonacci sequence for $h_k$. We will use two different settings to study transport, one will use an explicit boundary driving that will cause the system to converge to a time-independent nonequilibrium steady state (NESS), the other will be looking at a unitary evolution of a particular initial state. Both give consistent results with the same value of diffusion constant. We shall first briefly describe both settings and then focus on the results (Subsection~\ref{sec:results}).
\begin{figure}[ht!]
\centerline{\includegraphics[width=0.5\linewidth]{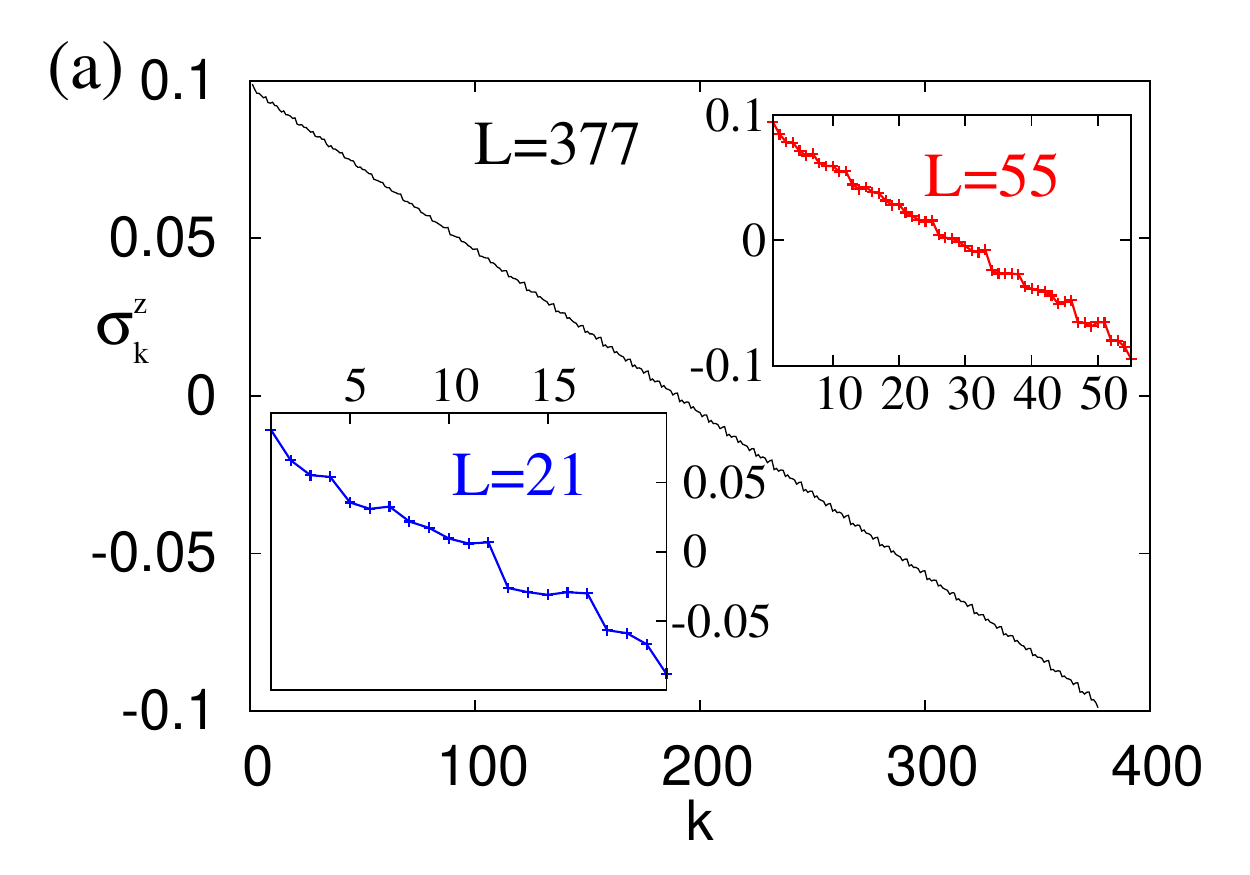}\includegraphics[width=0.5\linewidth]{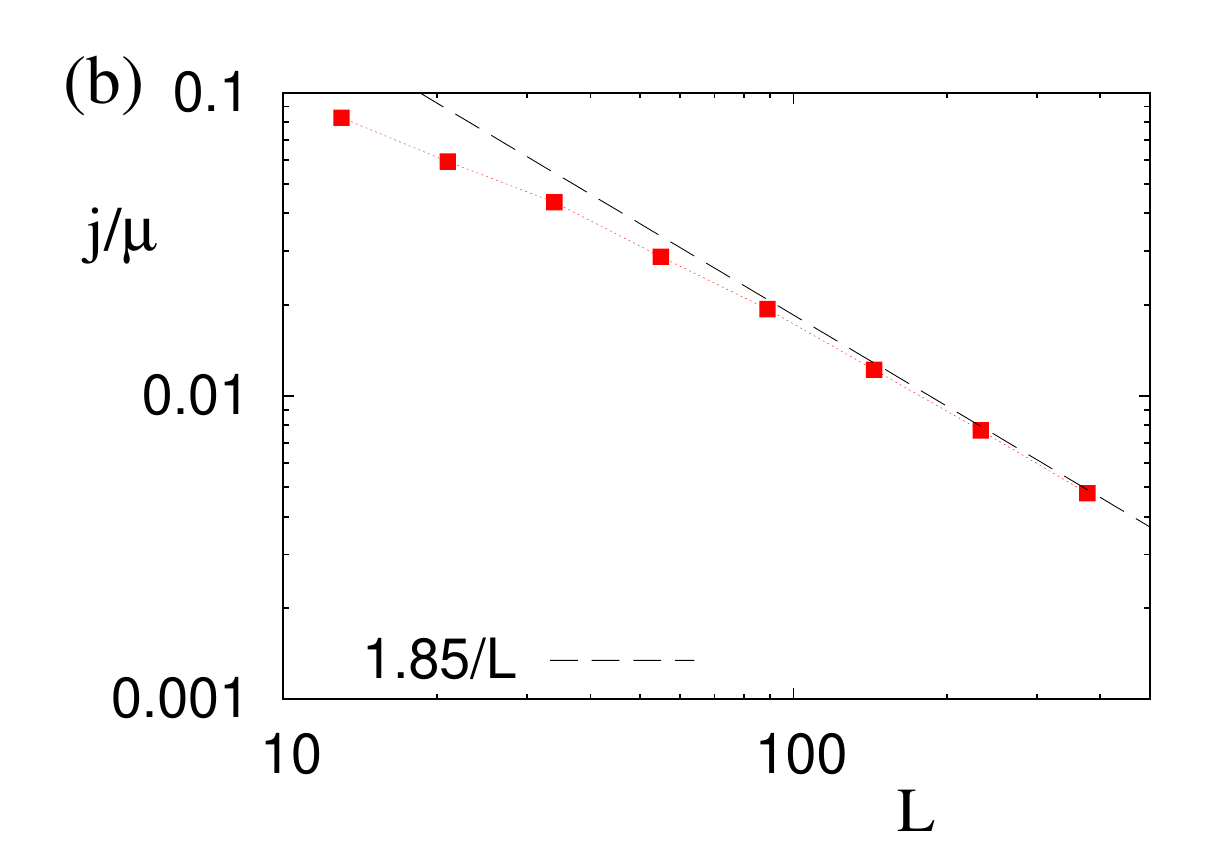}}
\centerline{\includegraphics[width=0.5\linewidth]{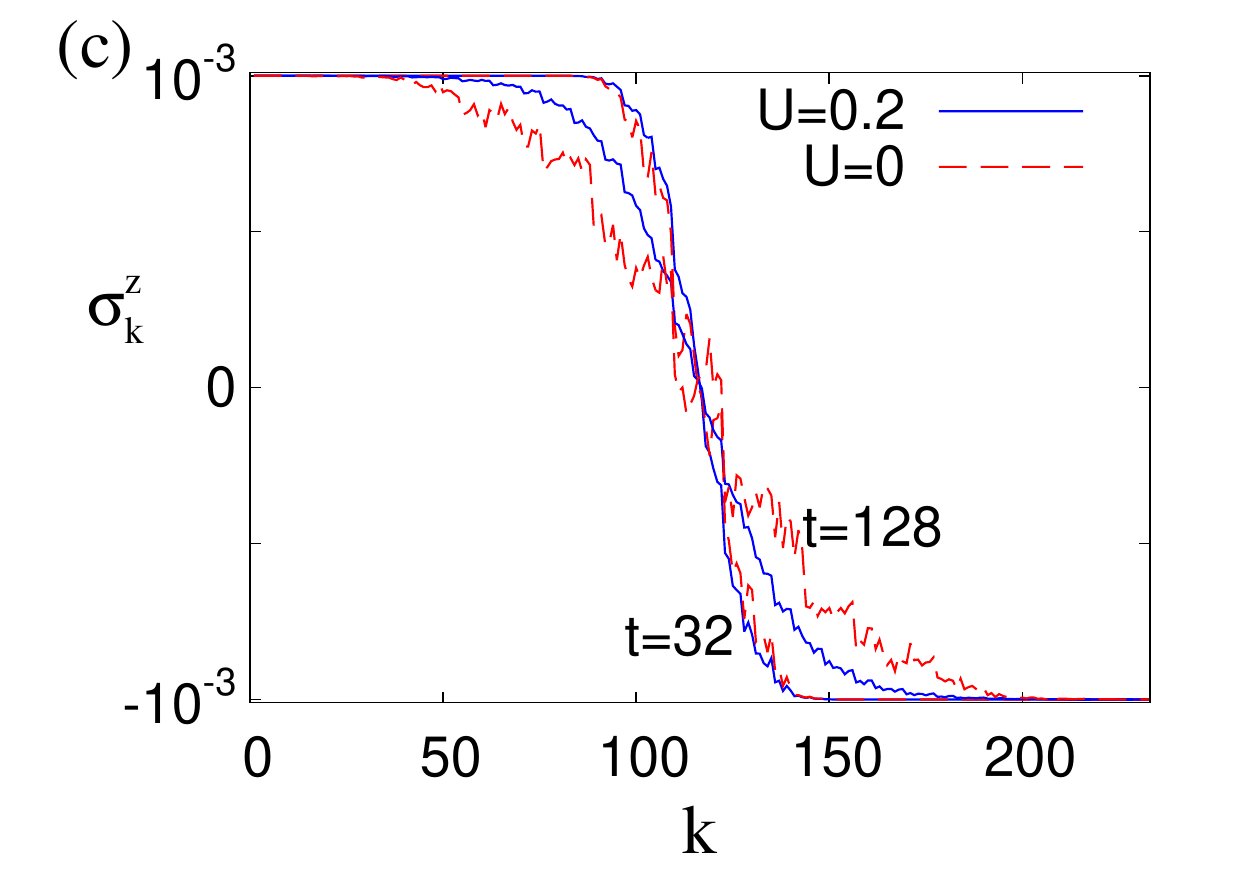}\includegraphics[width=0.5\linewidth]{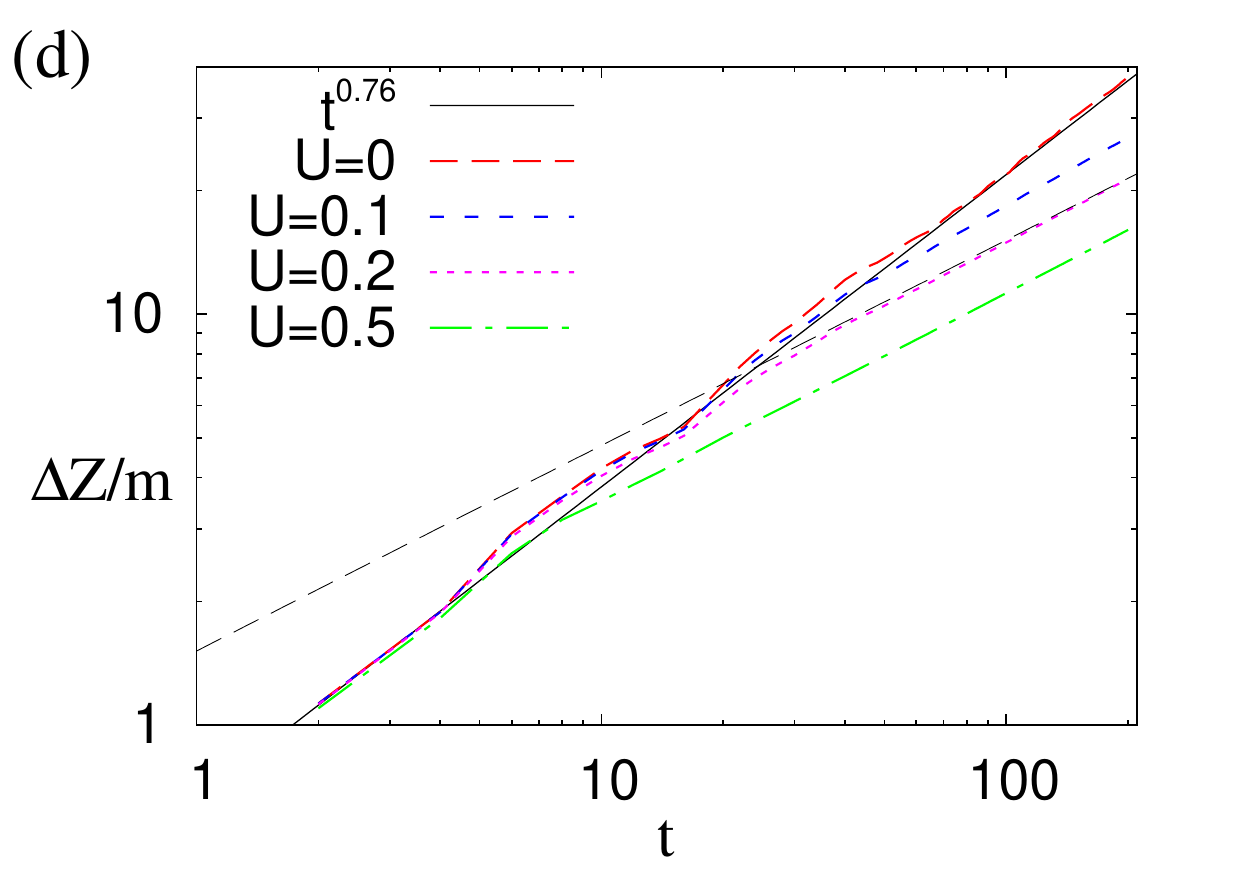}}
\caption{Top frames (a) and (b) demonstrate NESS physics for $U=0.2$; (a) shows magnetization profiles while (b) shows the current scaling, with the dashed line being the asymptotic diffusive $j/\mu=\Dn/L$ with $\Dn=1.85$. Bottom frames (c) and (d) are for the unitary evolution of a weakly polarized domain wall; (c) shows profiles at two times (dashed red curves are for $U=0$), while (d) is the transferred magnetization from the left to the right half of the chain (dashed black line fitting data for $U=0.2$ is diffusive $\sqrt{4\Du t/\pi}$ with $\Du=1.8$). All figures use $h=0.6$.}
\label{fig:demo}
\end{figure}

\subsection{Nonequilibrium steady state}

The simplest way to describe a nonequilibrium setting with an explicit driving that obeys all the rules of quantum mechanics (preserving positive semi-definiteness of $\rho$) is via the Lindblad~\cite{Lindblad1,Lindblad2} master equation,
\begin{equation}
{{\rm d}}\rho/{{\rm d}t}=\ii [ \rho,H ]+ {\cal L}^{\rm dis}(\rho).
\label{eq:Lin}
\end{equation}
There are two parts in the r.h.s. generator: one is the standard commutator that would, standing alone, generate unitary evolution, and the other is a so-called dissipator ${\cal L}^{\rm dis}$ that effectively describes a bath. The dissipator is written in terms Lindblad operators $L_k$ as ${\cal L}^{\rm dis}(\rho)=\frac{1}{4}\sum_k \left( [ L_k \rho,L_k^\dagger ]+[ L_k,\rho L_k^{\dagger} ] \right)$. Lindblad operators that we use are $L_1=\sqrt{1+\mu}\,\sigma^+_1, L_2= \sqrt{1-\mu}\, \sigma^-_1$ at the left end and $L_3 =  \sqrt{1-\mu}\,\sigma^+_L, L_4= \sqrt{1+\mu}\, \sigma^-_L$ at the right end, $\sigma^\pm=(\sigma^{\rm x} \pm {\rm i}\, \sigma^{\rm y})/2$. They represent magnetization reservoirs at infinite temperature and induce magnetization (spin) transport. 
Namely, the driving parameter $\mu$ essentially determines magnetization at the edges such that in long non-ballistic systems the boundary magnetization converges to $\ave{\sz{1}}=\mu$ at the left edge and 
to $\ave{\sz{L}}=-\mu$ at the right edge. 
We use a small $\mu=0.1$ so that all our results are in a linear response regime. Note that for $\mu=0$ the steady state would be $\rho_\infty \propto \mathbbm{1}$, 
i.e., an infinite temperature equilibrium state, and so small $\mu$ will results in a steady state that is at infinite temperature. 
Such a driving induces only magnetization flow (energy flow is zero). While justifying strong coupling that we use ($||L_k|| \sim 1$) on microscopic grounds is hard~\cite{breuerbook} in the thermodynamic limit (TDL) details of a boundary driving should not matter for bulk physics. This can indeed be rigorously shown~\cite{nessKubo} for diffusive systems.

After long time a time-dependent solution $\rho(t)$ of eq.~(\ref{eq:Lin}) converges to a nonequilibrium steady state (NESS), $\rho(t) \to \rho_\infty$, which is in our case unique. We find NESS $\rho_\infty$ by evolving $\rho(t)$ using time-dependent renormalization group method (tDMRG)~\cite{ulirev}. The method proceeds by writing expansion coefficients of $\rho(t)$ in a Pauli basis as a product of matrices, a so-called matrix product operator (MPO) ansatz, and then evolving it in time by Trotterization into small time steps of length $dt=0.2$ (we use a 4th order Trotter-Suzuki decomposition). Our adaptation for nonunitary evolution is described in Ref.~\onlinecite{NJP10}. Two parameters that determine the required computational effort are the relaxation time needed to converge to $\rho_\infty$ and the matrix product operator (MPO) bond dimension $\chi$. The relaxation time will increase with system size $L$, and similarly increase when transport gets slow. This makes the method work best for non-localized phases. How the required $\chi$ depends on physical properties is harder to say in advance; we typically find that one needs larger bonds for larger $h$ and/or when one approaches a possible subdiffusive transition. If one can afford to have bond sizes of several hundred~\cite{foot00}, and relaxation times $\sim 10^3$, this in some cases allows to study system sizes up to $L \sim 1000$ sites (see e.g. Ref.~\onlinecite{PRL16}), making the method in that regime by far the best one.

Once the NESS is obtained one can calculate the expectation values in the steady state. For the question of transport the most important ones are the expectation value of local magnetization $\ssz{k}=\frac{1}{2}\sz{k}$ and the associated spin current $j$, 
\begin{equation}
j:={\rm tr}[(\ssx{k}\ssy{k+1}-\ssy{k}\ssx{k+1})\rho_\infty],
\end{equation}
which is, due to continuity equation, independent of the site index $k$. Expectation value of $\ssz{k}$ in the NESS goes (in the TDL) from $+\mu/2$ to $-\mu/2$ across the chain so that for the case of diffusion one has the Fick's law
\begin{equation}
j \asymp D_{\rm NESS}\frac{\mu}{L}.
\end{equation}
One can get diffusion constant $D$ from the asymptotic slope of $j/\mu \asymp D_{\rm NESS}/L$. If one has anomalous transport the current would instead scale as
\begin{equation}
\frac{j}{\mu} \asymp \frac{1}{L^\gamma},
\label{eq:anom}
\end{equation}
with a scaling exponent $\gamma \neq 1$. $\gamma>1$ signifies subdiffusion while $\gamma<1$ would describe superdiffusion ($\gamma=0$ indicating ballistic transport).

The whole machinery is illustrated in the top Fig.~\ref{fig:demo}. We always calculate $\rho_\infty$ for several $\chi$, thereby checking the convergence as well as getting an estimate for the error due to finite $\chi$. In all plots we show data for the largest $\chi$, or, if the estimated errors are larger than about $\sim 1\%$, the extrapolated data. For details on the convergence with $\chi$ see Appendix~\ref{app:convergence}. In Fig.~\ref{fig:demo}(a) we can see that for small $h$ and $U$ and small system sizes $L$ profiles are not yet linear as one would expect for a diffusive system. However for larger systems, e.g. $L=377$, the asymptotic regime is reached with linear magnetization profile. This is also reflected in the scaling of current $j$ with system size, Fig.~\ref{fig:demo}(b). For too small systems $j(L)$ does not yet ``feel'' the full interacting dynamics and $j(L)$ falls with $L$ in a slower superdiffusive fashion, being though just a remnant of the superdiffusive noninteracting physics (at $h=0.6$). However, at larger $L$ the true asymptotic transport is reached with diffusive $j \sim 1/L$ scaling.

\subsection{Unitary evolution}

Another way to probe transport is by looking at the spreading on inhomogeneous initial states. For instance, starting with an initial wave-packet that has a non-stationary Gaussian profile of magnetization one could look at how the width $\Delta x$ of the packet grows with time,
\begin{equation}
\Delta x \sim t^\beta,
\label{eq:alpha}
\end{equation}
introducing a transport scaling exponent $\beta$. If one has a single scaling exponent then simple dimensional analysis gives a relation between $\beta$ of the unitary evolution and $\gamma$ of the NESS, namely
\begin{equation}
\beta=\frac{1}{\gamma+1}.
\label{eq:ag}
\end{equation}

To assess the asymptotic transport one needs to simulate evolution up to as long times as possible. For that one again uses the tDMRG method on a matrix product ansatz. A decisive quantity is how fast the ``entanglement'' grows with time, or equivalently, how far in time one can go with a given maximal $\chi$. While one could evolve a pure state, it turns out that doing unitary evolution on an ensemble of states, that is on a density operator $\rho$, can be (structurally) more stable, allowing to simulate longer times. A good choice of an initial state for studying magnetization transport is a weakly polarized domain wall~\cite{natcomm},
\begin{equation}
\rho(0) \propto \prod_{k=1}^{L/2}(\mathbbm{1}+m\, \sz{k}) \otimes \prod_{k=L/2+1}^{L}(\mathbbm{1}-m\, \sz{k}).
\label{eq:dw}
\end{equation}
Transport type can then be assessed by e.g. looking at how magnetization transported about the mid-point of the chain grows with time. That is,
\begin{equation}
\Delta Z:=m\frac{L}{2}-\sum_{k=1}^{L/2} \ave{\sz{k}(t)},\quad \Delta Z \asymp t^\beta.
\end{equation}
In case of diffusion ($\beta=\frac{1}{2}$) magnetization profile will converge to a shape given by the error function
\begin{equation}
\label{eq:err}
\ave{\sz{k}(t)} \asymp -m\, {\rm erf} \left( \frac{k-L/2}{\sqrt{4\Du t}}\right),
\end{equation}
while the transferred magnetization will grow as
\begin{equation}
\frac{\Delta Z}{m} \asymp \sqrt{\frac{4 D_{\rm unit} t}{\pi}}.
\label{eq:dZ}
\end{equation}

In all our simulations we use small $m=10^{-3}$ meaning that we are again in a linear response regime. An example of such a simulation is shown in the bottom Fig.~\ref{fig:demo}. 
One obvious observation from Fig.~\ref{fig:demo}(d) is that the interacting case starts to differ from a noninteracting one only after times larger than $\sim 1/U$. 
After that a further transient time is required to reach the asymptotic transport. A similar conclusion can be reached observing magnetization profiles in Fig.~\ref{fig:demo}(c). At short $t=32$ both interacting and noninteracting profiles look similar and very ``noisy'' due to (non-interacting) multi-fractality. However, at longer time $t=128$ one can see that the interacting dynamics slows down and the profile becomes much smoother -- both being manifestations of the asymptotic diffusive dynamics.

We again stress, see also e.g. Ref.~\onlinecite{PRL16}, that if the integrability breaking perturbation is small, be it a small $h$ or a small $U$, large times (large systems in the NESS setting) are needed in order to see the true transport type~\cite{foot0}. Ignoring that can lead to incorrect results, an example being Ref.~\onlinecite{barlev17} where too short times are used rendering most of their claims false~\cite{foot1}.

\subsection{Diffusion for small $U$}
\label{sec:results}

At small $U$ and for a range of potential strengths $h \le 3$ that we are able to reliably simulate we always find diffusive magnetization transport (at high temperature and half-filling that we study). The diffusion constant agrees between the NESS and unitary evolutions, see Table~\ref{tab1} . 
Different methods aiming at the same physical quantity should of course agree, but we note that empirically that is far from being an established fact. 
For instance, for random potential different publications often report even different transport types (see e.g. comparison in Fig.~1 of Ref.~\onlinecite{luitz17}). 
The agreement that we find is therefore nontrivial and gives a further weight to our results. For the boundary driven Lindblad equation that we use one can in fact derive a NESS version of Kubo formula and show rigorously~\cite{nessKubo} that if the dynamics is diffusive the two settings give exactly the same diffusion constant $D$, with a finite-size correction being of order $\sim 1/L$.
\begin{table}[ht!]
\begin{center}
\begin{tabular}{l||c|c|c|c}
 & \multicolumn{4}{c}{$4D_{\rm NESS}$ [$4D_{\rm unit}$]}\\
$U$ &  $h=0.6$ & $h=1.0$ & $h=2.0$ & $h=3.0$ \\
\hline
0.05 & & $3.6_{\pm 2}$ & $0.23_{\pm 1}$ & $0.065_{\pm 7}$ \\
0.1 & $14_{\pm 1}$ & $2.75_{\pm 5}$ [$2.68_{\pm 3}$] & $0.27_{\pm 1}$ & $0.070_{\pm 8}$\\
0.2 & $7.4_{\pm 1}$ [$7.2_{\pm 2}$] & $2.03_{\pm 4}$ [$2.01_{\pm 2}$] & $0.24_{\pm 1}$ [$0.24_{\pm 1}$] & $0.047_{\pm 4}$\\
0.5 & $4.1_{\pm 1}$ [$4.0_{\pm 2}$] & $1.31_{\pm 3}$ & subdiffusion & \\
\end{tabular}
\caption{Values of the diffusion constant (and last-digit's uncertainty, e.g., $3.6_{\pm 2}=3.6 \pm 0.2$) obtained from NESS simulations and from unitary evolution of a weakly polarized domain wall (numbers in square brackets). In the table we report the values of $4\cdot D$, which would be the diffusion constant if we would use Pauli operators, like $\sx{k}\sx{k+1}$, instead of spin-1/2 ones, like $\ssx{k}\ssx{k+1}$.}
\label{tab1}
\end{center}
\end{table}

\begin{figure}[ht!]
\centerline{\includegraphics[width=0.85\linewidth]{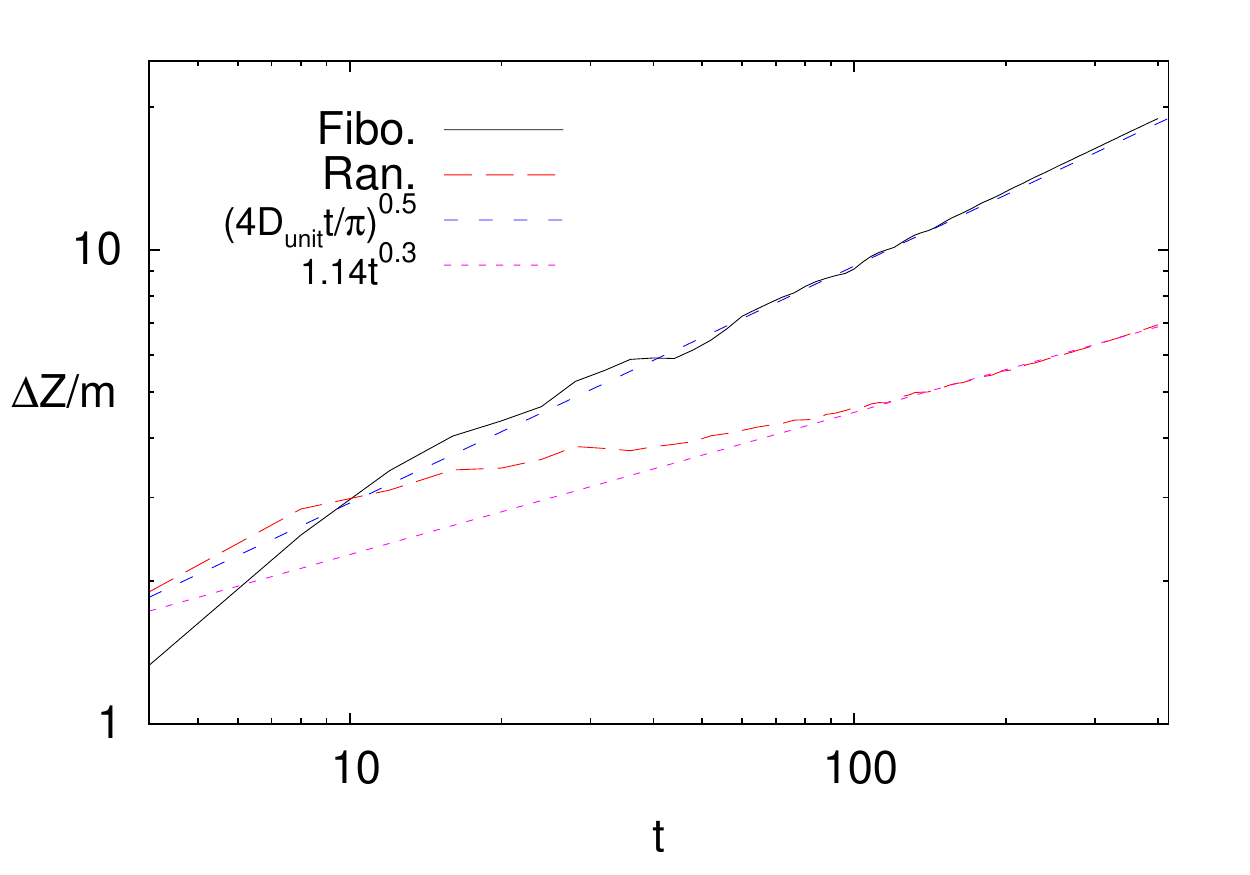}}
\caption{Unitary evolution of a weakly polarized domain wall for the Fibonacci potential (black full curve, average over 5 potential instances) and random binary potential (red dashed curve, average over 20 random instances) for the same parameters $h=1$, $U=0.1$ and $L=144$. 
Transport in the quasiperiodic Fibonacci case is diffusive ($\Du=0.67$), while it is subdiffusive for the random case.}
\label{fig:rand}
\end{figure}
Obtaining diffusion for an arbitrarily small $U$ is surprising. Recall that in the free model ($U=0$) the transport type continuously varies with $h$ and therefore an arbitrary non-interacting $\beta$ immediately breaks down to a diffusive $\beta=1/2$ for nonzero $U$. We have to stress that this is very different than for a random potential. There the breakdown happens continuously, going from Anderson localization for $U=0$ through a subdiffusive regime for small nonzero $U$. That is, if instead of a Fibonacci sequence of potential values, for which one has diffusion, one takes a random choice $h_j=\pm h$ at each site, one instead gets subdiffusion, see Fig.~\ref{fig:rand}. Long-range correlations of the Fibonacci potential are crucial for the observed diffusive transport. Such a discontinuous change seems in fact to be a common property of quasiperiodic potentials, it is for instance also observed~\cite{pnas18} for a cosine potential in the AAH model, where it could not be explained by simple perturbation theory. 

\begin{figure}[t!]
\centerline{\includegraphics[width=0.85\linewidth]{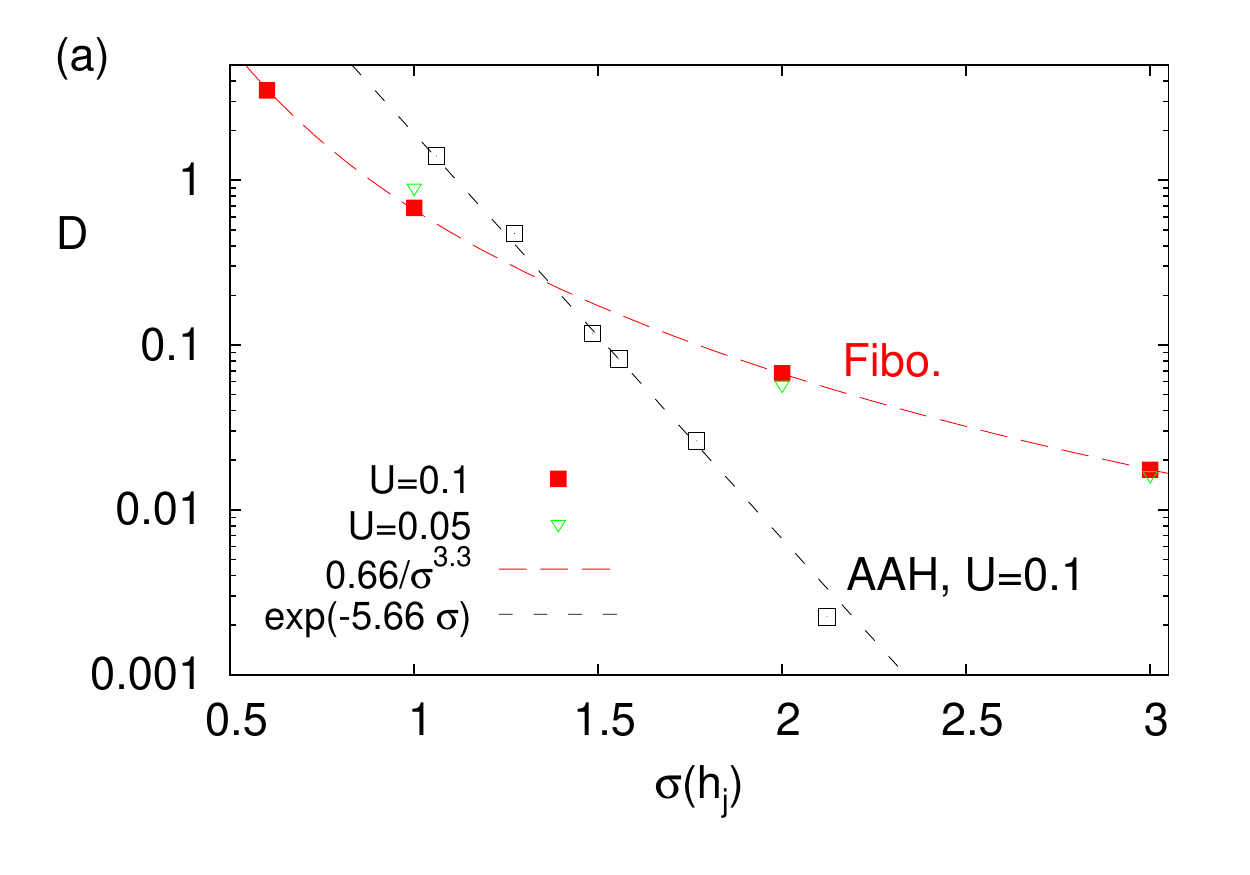}}
\centerline{\hskip3mm\includegraphics[width=0.81\linewidth]{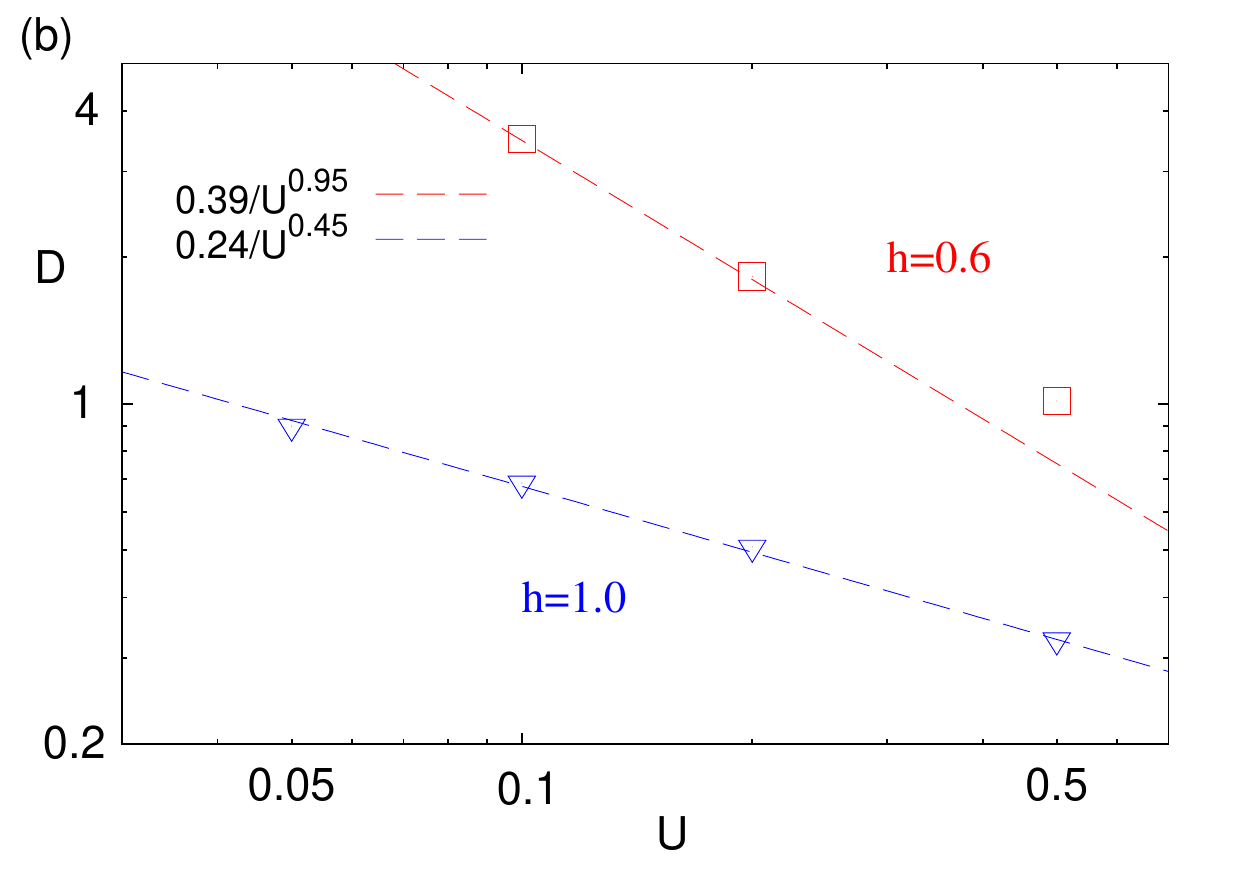}}
\caption{Dependence of the diffusion constant on $h$ (top) and $U$ (bottom). (a) At fixed $U$, the diffusion constant has a power law dependence on potential amplitude for the Fibonacci model (red and green points), while it has an exponential dependence for the Aubry-Andre-Harper 
(AAH) model that has a cosine potential (black squares, data from Ref. \onlinecite{pnas18}). Everything is plotted as a function of potential variance, being $\sigma(h_j)=h$ for the Fibonacci model and $\sigma(h_j)=h/\sqrt{2}$ for the AAH model. (b) For the Fibonacci model diffusion constant scales as $D \sim 1/U^\nu$ for small $U$ and fixed $h$.
}
\label{fig:hU}
\end{figure}
A simplistic picture could use Fermi's golden rule on perturbative interaction to argue for the diffusion. This would predict the scattering rate due to small interaction to scale as $U^2$, which should be in turn reflected in the scaling of $D$ for small $U$. 
Simply using the scaling $x \sim t^\beta$ between space and time would give the scattering length diverging as $\sim 1/U^{2\beta}$ (note that we are neglecting any fractality of matrix elements, which changes as a function of $h$). Using the scaling ansatz~\cite{PRL16} $j \sim U^{2\beta-\nu} f(L\, U^{2\beta})$ for the NESS current, where the scaling function must have the asymptotics $f(x \gg 1)\sim 1/x$ due to diffusion for large $L$, and $f(x\ll 1) \sim 1/x^\gamma$ due to an anomalous non-interacting scaling $j \sim 1/L^{\gamma}$ for $U \to 0$, gives that the diffusion constant diverges as $D \sim 1/U^\nu$ with $\nu=2\beta(1-\gamma)$. Using Eq.~(\ref{eq:ag}) one gets $\nu=4(\beta-\frac{1}{2})$. In Fig.~\ref{fig:hU}(b) we can see that for small $U$ diffusion constant indeed diverges as $D \sim 1/U^\nu$ with the scaling exponent $\nu$ increasing for decreasing $h$, and being roughly consistent with the relation $\nu=4(\beta-\frac{1}{2})$. To really confirm this relation however more data would be required. For larger $h$, e.g. $h\ge 2$, we would presumably need still smaller $U$ to see a clear power-law dependence with $U$ (with a negative $\nu$). 
Namely, for potential amplitudes where the noninteracting case is subdiffusive we observe a nonmonotonous dependence of $D$ on $U$, with a maximum being reached at a fairly small $U \approx 0.1$ (see Table~\ref{tab1} as well as Fig.~\ref{fig:maxU}). We would thus necessitate very small $U$ to reveal small-$U$ behavior, in turn requiring very large systems. As a function of $h$ at fixed $U$ (Fig.~\ref{fig:hU}(a)) one has a power-law dependence in the available range of $h$, $D \sim 1/h^\xi$, with the power $\xi \approx 3.3$ for $U=0.1$ (and slightly increasing for smaller $U$). This must be contrasted with the AAH models where the dependence on $h$ is exponential.

Comparing the cases of random, cosine-quasiperiodic, and Fibonacci quasiperiodic potential one can say that transport is the fastest in the Fibonacci model, then in the AAH, while it is the slowest for random potential. All quasiperiodic potentials are faster than random because of long-range correlations in the potential (in a sense, a quasiperiodic potential is ``almost'' periodic, in which case one would have ballistic transport). That the Fibonacci is faster than the AAH can on the other hand be ascribed to a broad momentum-spectrum of the potential~\cite{monthus17}.

\subsection{Large interaction}

So far we have found only diffusive transport. An important question is can one perhaps get a subdiffusive transport at larger $U$? After all, one of the reasons why quasiperiodic potential is interesting is to clarify the influence of rare-regions that are argued to be responsible for subdiffusion in random potential~\cite{agarwal15,sarang16,Adam18}, and, of course, also from a fundamental desire to understand under which circumstances can one get an anomalous transport. 

Let us say right away that the question of larger $U$ is very important and demands a careful separate study -- in this work we predominantly focus on small interactions. 
Namely, at larger $U$ where one approaches a possible phase transition into subdiffusion, numerics typically gets harder (first, relaxation gets longer because the transport gets slower, and second, one also typically needs fairly large $\chi$). Here we therefore report only on a situation for two sets of parameters.

\begin{figure}[ht!]
\centerline{\includegraphics[width=0.85\linewidth]{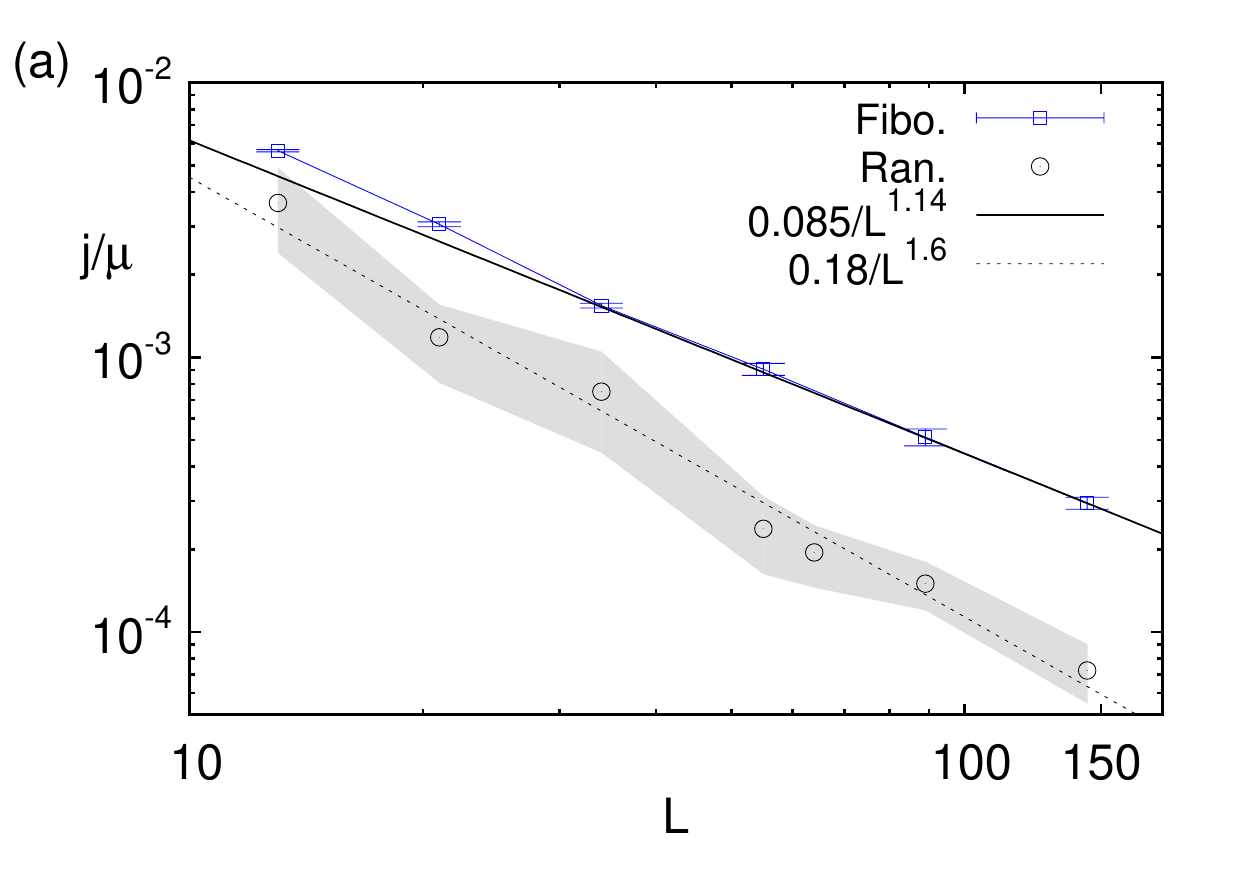}}
\centerline{\includegraphics[width=0.85\linewidth]{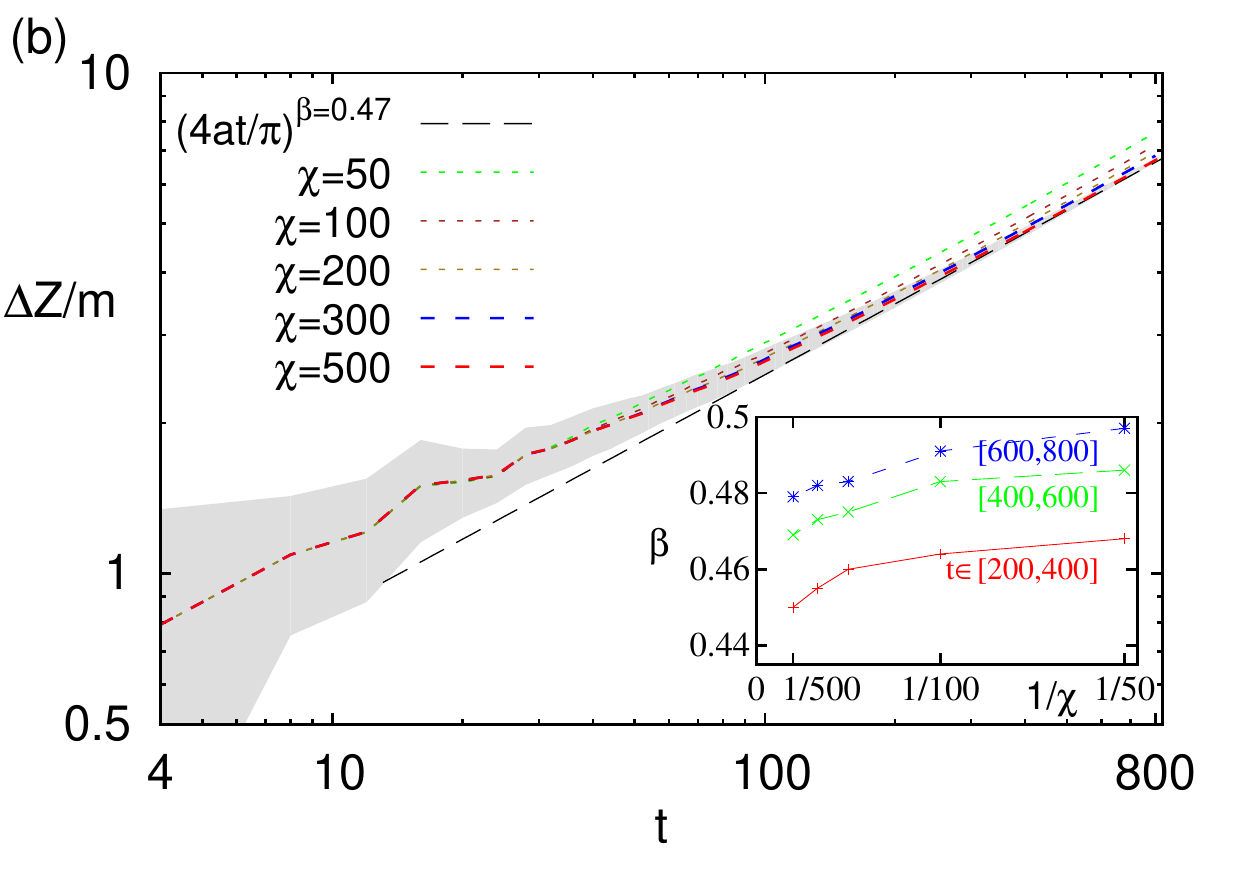}}
\caption{Subdiffusive transport for $U=0.5$ and $h=2$. (a) NESS setting and the Fibonacci model (blue squares, one sample, error bars show the estimated error due to finite bond size $\chi\approx 300$), and for comparison the case of random disorder with $h_j=\pm h$ (black circles). Gray shaded region is a rough estimate of the variance over realizations (obtained from only 2-3 samples). (b) Unitary dynamics of a weakly polarized domain wall for the Fibonacci model, again showing subdiffusion with $\beta \approx 0.47$ and $a=0.055$, being roughly consistent with $\gamma=1.14$ from (a). Average over 4 Fibonacci sequences with $L=144$ (there are no boundary effects for the shown times), gray shading is the standard deviation, indicating self-averaging for large $t$. The inset shows convergence of $\beta$ with $\chi \in [50,500]$ for different fitting windows.}
\label{fig:sub}
\end{figure}
Data for $U=0.5$ and $h=2$ is shown in Fig.~\ref{fig:sub}. Numerics is quite demanding, for instance, despite using $\chi=300$ in NESS simulations we could achieve only about $5 \%$ accuracy in $j$ for the 
largest $L=144$. Data in Fig.~\ref{fig:sub}(a) is best fitted by $j \sim 1/L^{1.14}$. 
Even taking into account the mentioned $\approx 5 \%$ error we can with statistical significance say that the exponent $\gamma$ is larger than $1$ and one therefore has subdiffusion. 
Observe also that for the same parameters, using random binary potential (i.e., just reordering locations of $+h$ and $-h$ potential from the Fibonacci to a random one) results in much stronger subdiffusion. 
We remind that as far as localization is concerned, a binary disorder potential behaves similarly~\cite{sirker17,delande17} 
as the more frequently studied one with a box distribution. 
Further confidence that we are indeed seeing subdiffusion is provided by unitary evolution of a weakly polarized domain wall shown in Fig.~\ref{fig:sub}(b). Here one again needs large $\chi$ and one in particular needs to increase $\chi$ for larger times, similarly as in NESS simulations where one has to use larger $\chi$ for larger $L$ if one wants to keep the accuracy constant. Having too small $\chi$ will tend to push dynamics towards a (fake) diffusion, 
see e.g. data for $\chi=50-100$ in Fig.~\ref{fig:sub}(b). Too large truncation due to small $\chi$ has a similar effects as classical noise, e.g. dephasing, 
which will in the TDL always cause diffusion regardless of whether one has interaction~\cite{NJP10,ann17} or not~\cite{XXdeph13}. The best fitted exponent $\beta \approx 0.47$ is consistent via eq.(\ref{eq:ag}) with the NESS $1/(1+\gamma) \approx 0.467$. We also observe that for times smaller than about $t \approx 300$ the dynamics is not yet the asymptotic one -- the exponent has clearly not yet converged -- while at larger times we converge to the same $\beta<0.5$ regardless of time (we of course can not exclude that at still larger times $t \gg 10^3$ and $L \gg 144$ one would eventually end up with diffusion, however for $L\le 144$ and $t<800$ we find no indications of that).

Observing subdiffusion in a quasiperiodic potential clearly shows that ``rare-regions'' can not serve as a universal ``explanation'' of subdiffusion. 
That this is so should be clear also from the observed subdiffusion in a non-interacting model for $h>h_{\rm diff}\approx 1.6$!

\begin{figure}[t!]
\centerline{\includegraphics[width=0.8\linewidth]{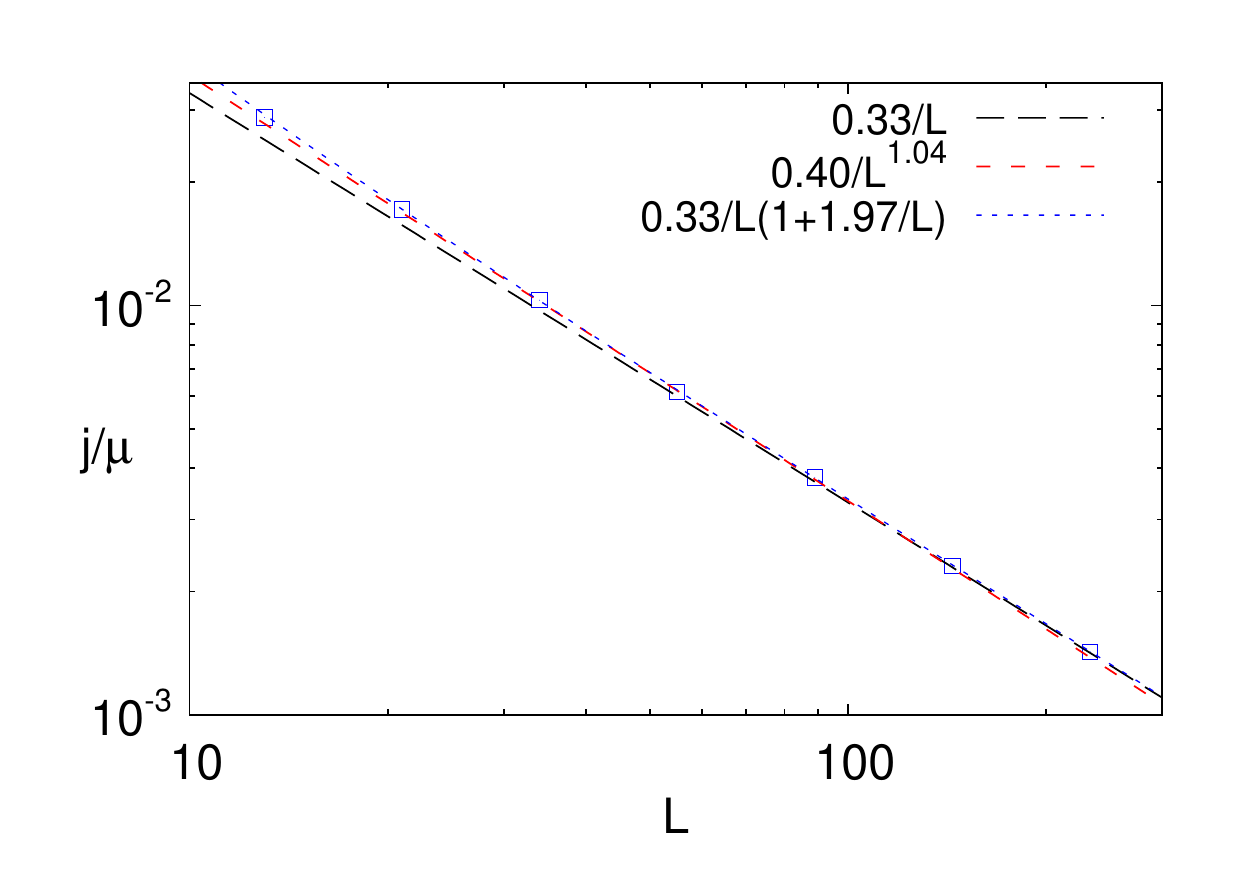}}
\caption{Diffusive transport for $U=0.5$ and $h=1$.}
\label{fig:U05h1}
\end{figure}

If, on the other hand, one takes smaller $h=1$ and the same interaction $U=0.5$, Fig.~\ref{fig:U05h1}, one instead of subdiffusion still sees diffusion. Because it is hard to distinguish marginal subdiffusion with $\gamma$ being close to $1$ from true diffusion $\gamma=1$, we fit to the NESS current data two curves each having two free parameters: one is $j \sim a/L+b/L^2$ (theory for diffusion predicts~\cite{nessKubo} finite-size $1/L^2$ correction), while the other is $j \sim a/L^b$. One can see (Fig.~\ref{fig:U05h1}) that the first one fits better and therefore we deem to still have diffusion (phase transition could though be close to the point $U=0.5$, $h=1$).
For $h=1$ and $U=1$ the data we gathered for $L \le 233$ (not shown) also indicated the asymptotic diffusion with $4\Dn \approx 0.96$. At smaller $L\le 50$ a weak subdiffusion $\gamma \approx 1.1-1.2$ is observed so one needs large systems to observe the asymptotics (entanglement growth with weakly ``subdifusive'' scaling exponent $1/z \approx 0.82$ was observed for $U=1, h=1$ and times $t <20$ in Ref.~\onlinecite{Alet18}).

\section{Mean-field failure}

\subsection{Fully polarized domain wall}

\begin{figure}[ht!]
\centerline{\includegraphics[width=0.8\linewidth]{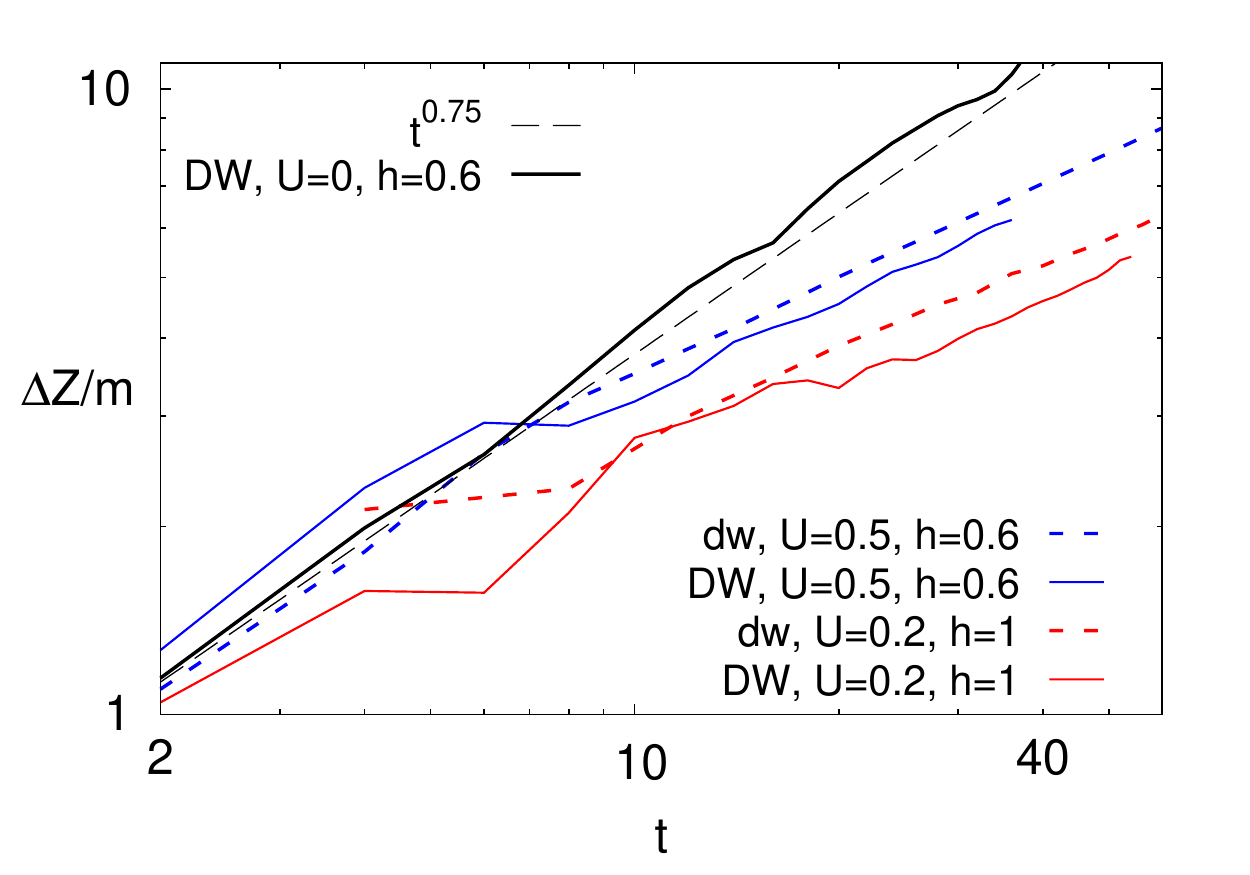}}
\caption{Comparison between unitary evolution of a mixed state weakly polarized domain wall (dw, red and blue dotted curves) and a pure state fully polarized domain wall (DW, red and blue full curves). Also shown is a noninteracting case for $h=0.6$ where $\beta \approx 0.75$ (full black curve). Mixed state data (label dw) uses $L=233$ and $\chi=200$, pure state (label DW) $L=89$ and $\chi=700$.}
\label{fig:DWcompare}
\end{figure}

All unitary data on previous pages was for a weakly polarized mixed state domain wall with polarization $m= 10^{-3}$. While for a system without potential a fully polarized domain wall (DW) pure state
\begin{equation}
  \ket{{\rm DW}}=\ket{\uparrow \cdots \uparrow \downarrow \cdots \downarrow},
\label{eq:DW}
\end{equation}
is a special state and can not be used to infer (generic) transport, this is not so for a model with nonzero $h_j$ like our Fibonacci chain. We therefore check if the dynamics of the DW leads to the same transport as that of a weakly polarized mixed state domain wall (\ref{eq:dw}), or of a driven NESS. Using pure state tDMRG we evolve the state and calculate the transfered magnetization $\Delta Z$ (\ref{eq:dZ}). In Fig.~\ref{fig:DWcompare} we compare the results with those for a weakly polarized domain wall. While we can not reach times longer than about $t \approx 40$ for pure-state evolution (despite having $\chi \approx 700$) it seems that the growth of $\Delta Z(t)$ is similarly diffusive as for a weakly polarized domain wall.

\subsection{Mean field dynamics}

Let us now try to describe the dynamics of a fully polarized domain wall using a mean field treatment of the interaction. Standard mean field treatment of a product of two operators $A B$ neglects the term quadratic in fluctuations, $(A-\ave{A})(B-\ave{B})$, leading to a mean field replacement $AB \to A\ave{B}+\ave{A}B$. Doing that on the interaction $U \sz{k}\sz{k+1}$ gives $U(\sz{k}\ave{\sz{k+1}}+\ave{\sz{k}}\sz{k+1})$, rendering the model noninteracting with a modified on-site potential $h_k \to h_k + U (z_{k-1}+z_{k+1})$, where we denote $z_k:=\bracket{\psi(t)}{\sz{k}}{\psi_k(t)}$. A possible ``explanation'' of the observed diffusive dynamics that suggests itself is the following. We have an XX chain with a Fibonacci potential and fluctuating magnetic fields. Provided the fluctuating magnetic field is Gaussian and uncorrelated in time it is equivalent to a so-called dephasing Lindblad equation. That in turn, no matter how small is the dephasing, in the thermodynamic limit always leads to diffusion~\cite{XXdeph13}. However, as we show, such a ``plausible'' explanation of the observed diffusion, is, in fact, wrong.

We shall show that by studying the mean field equations of motion. Although they fail to correctly describe dynamics in a quasiperiodic potential, an interesting open question is whether they can describe dynamics in some other situations, for instance, in a random potential. Let us first write down equations of motion for expectation values of observables in a noninteracting model. Expanding density operator in terms of products of Pauli operators,
\begin{equation}
  \rho \propto \mathbbm{1}+\sum_k z_k \sz{k}+ \sum_k (xy-yx)_k (\sx{k}\sy{k+1}-\sy{k}\sx{k+1})+\cdots,
\label{eq:rho}
\end{equation}
where e.g. $z_k$ and $(xy-yx)_k$ are abbreviations for corresponding expansion coefficients. Von Neumann equation of motion for $\rho(t)$ gives equations for the expansion coefficients. Using Jordan-Wigner transformation all 2-point expectations of fermionic operators form a closed set of equations. In spin language all those expectation values can be compactly encoded in a Hermitian correlation matrix $C_{i,k}$, with the off-diagonal matrix elements being $C_{i,k}:=(xz\cdots zx+yz\cdots zy)_i+\ii (xz\cdots zy-yz\cdots zx)_i$, i.e., coefficients in front of $(\sx{i}\sz{i+1}\cdots \sz{k-1}\sx{k}+\sy{i}\sz{i+1}\cdots \sz{k-1}\sy{k})$ and $(\sx{i}\sz{i+1}\cdots \sz{k-1}\sy{k}-\sy{i}\sz{i+1}\cdots \sz{k-1}\sx{k})$ in $\rho$ (\ref{eq:rho}), while the diagonal is $C_{i,i}:=-z_i$. For instance, $C_{j,j+1}$ give coefficients in front of magnetization currents and the kinetic term (hopping), the expectation value of the current being ${\rm tr} [(\ssx{k}\ssy{k+1}-\ssy{k}\ssx{k+1})\rho ]=\frac{1}{2}{\rm Im}(C_{k,k+1})$). Equations of motion can now be compactly written as a matrix equation
\begin{equation}
  \frac{dC}{dt}=\frac{\ii}{2} [ C,{\cal H} ],\quad {\cal H}={\eta}-{\cal J},
  \label{eq:C}
\end{equation}
where ${\eta }$ comes from potential and is a diagonal matrix with elements ${\eta }_{k,k}=h_k$, while ${\cal J}$ represents hopping and has nonzero elements only along the two next-diagonals, ${\cal J}_{k,k\pm 1}={\cal J}_{k\pm 1,k}=1$. Note that similar equations, only with a non-Hermitian ${\cal H}$, govern also Lindblad dynamics~\cite{VarmaZnidaricPRE}.
For a fully polarized domain wall (\ref{eq:DW}) the initial condition is $C(0)=-{\rm diag}(1,\ldots,1,-1,\ldots,-1)$. \par
One can obtain different mean field descriptions depending on fluctuations of which operators are neglected. For our concrete interaction one can for instance work directly with $\sz{k}$, or, one can alternatively express $\sz{k}$ in terms of raising and lowering operators, like $\m{k}\p{k}=(\mathbbm{1}-\sz{k})/2$ and $\p{k}\m{k}=(\mathbbm{1}+\sz{k})/2$, and neglect fluctuations of those. Denoting for later reference the two mean field variants by (a) and (b), we have:
\begin{enumerate}[(a)]
\item $\sz{k}\sz{k+1} \to \ave{\sz{k}}\sz{k+1}+\sz{k}\ave{\sz{k+1}}$, amounting to
  \begin{equation}
    h_k \to h_k + U(z_{k-1}+z_{k+1}).
    \label{eq:mf1}
    \end{equation}
\item writing $\sz{k}\sz{k+1}=(\p{k}\m{k}-\m{k}\p{k})(\p{k+1}\m{k+1}-\m{k+1}\p{k+1})$, and doing mean-field on products of raising and lowering operators, results in
  \begin{eqnarray}
    \sz{k}\sz{k+1}\to && \frac{1}{2}(z_k \sz{k+1}+\sz{k} z_{k+1})-2 \m{k}\p{k+1}\ave{\p{k}\m{k+1}}-\nonumber \\
    &&-2\p{k}\m{k+1}\ave{\m{k}\p{k+1}},
    \label{eq:mf20}
  \end{eqnarray}
  plus a global constant. Noting that $\ave{\m{k}\p{k+1}}=C_{k,k+1}/2$ we get a mean field replacement rule for the potential and the hopping,
  \begin{eqnarray}
        \label{eq:mf2}
    h_k &\to& h_k+\frac{U}{2}(z_{k-1}+z_{k+1}),\\
    {\cal J}_{k,k+1} &\to& 1-\frac{U}{2} C_{k,k+1}, \quad {\cal J}_{k+1,k} \to 1-\frac{U}{2} C_{k+1,k}.\nonumber
    \end{eqnarray}
\end{enumerate}

In version (a) the mean field equations (\ref{eq:C}) therefore get an additional term $-\ii\frac{U}{2}[C,C_{\rm d}]$, with the diagonal $[C_d]_{k,k}:=C_{k-1,k-1}+C_{k+1,k+1}$, on the right-hand-side. In case (b) one has a similar terms plus an additional $\ii\frac{U}{4}[C,C_{\rm o}]$, where $C_{\rm o}$ is equal to two 1st off-diagonals of $C$.

\begin{figure}[t!]
\includegraphics[width=0.8\linewidth]{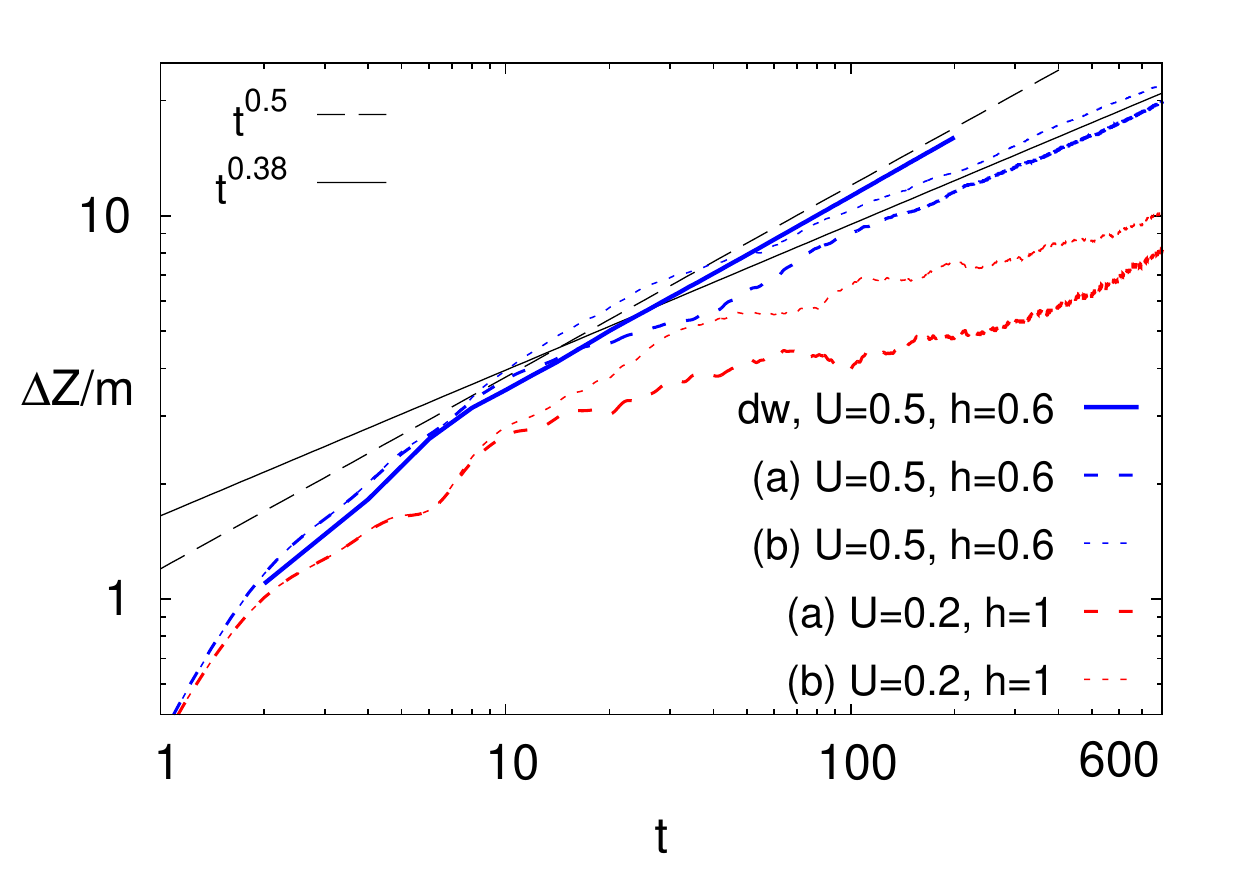}
\caption{Comparing two mean field variants, listed under (a) and (b) in the text, for a fully polarized domain wall in the Fibonacci chain. 
For comparison, we also show exact diffusive dynamics of a weakly polarized domain wall (labeled by dw) for $U=0.5$, $h=0.6$. 
Mean field asymptotically always incorrectly leads to subdiffusion. All data is for $L=233$ such that there are no finite-size effects.}
\label{fig:DWmf}
\end{figure}
\begin{figure}[t!]
\includegraphics[width=0.8\linewidth]{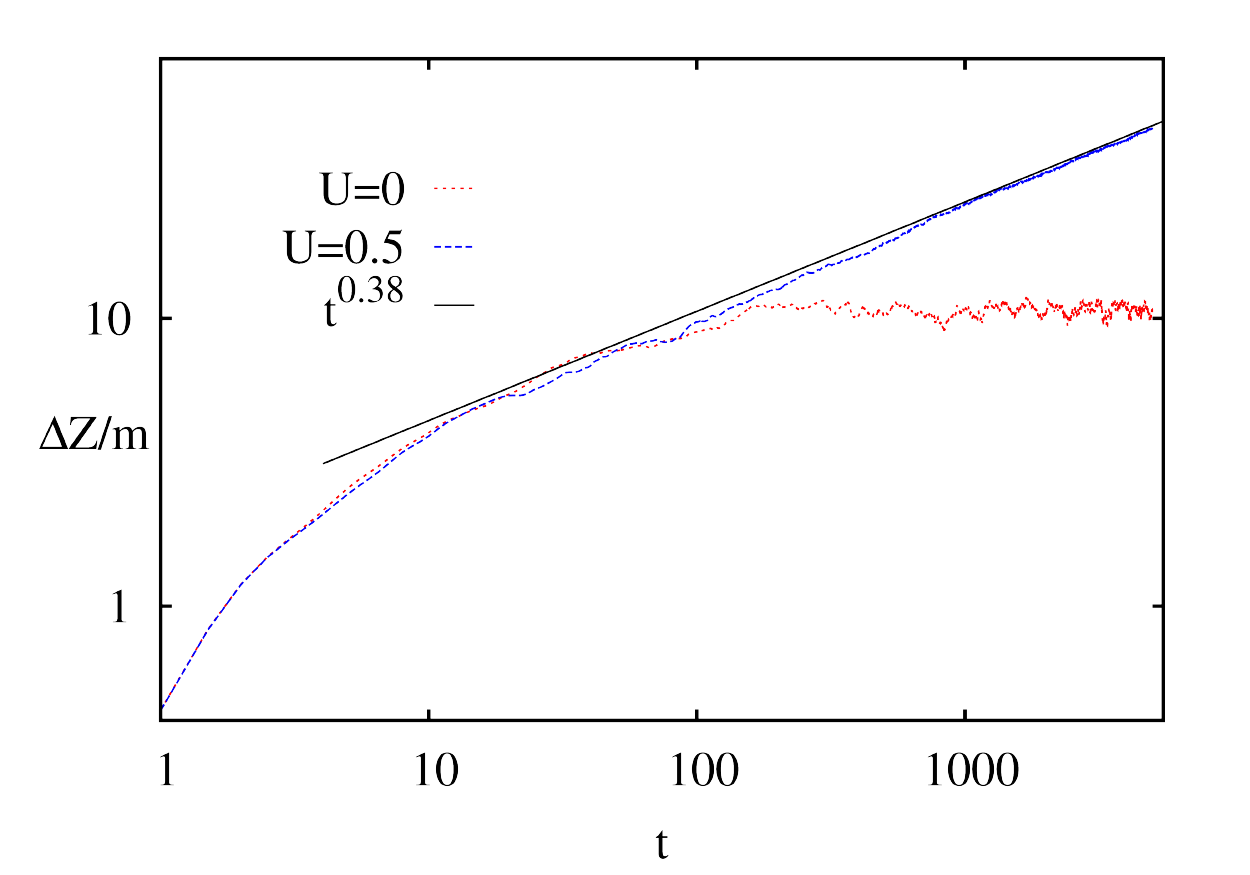}
\caption{Mean field protocol (a) for $L=233$ and $h=0.6$ for a binary-disordered chain. Upon introducing interactions, the erstwhile localized system starts transporting spin fluctuations; 
for the present case $U=0.5$ the transport is subdiffusive as indicated by the straight line fit.}
\label{fig:DWmf-bin}
\end{figure}

In both cases equations become nonlinear and their behavior has to be analysed numerically. We did that by integrating them, starting with a DW initial condition, and averaging over all $L+1$ different Fibonacci sequences. Results are shown in Fig.~\ref{fig:DWmf}. We can see that while the mean field description might look promising at short times (e.g., dashed mean-field curves might seem to follow exact diffusive dynamics up to $t \approx 20$), at longer times mean-field is subdiffusive rather than diffusive. 
We therefore must conclude that the mean field treatment can not explain the observed diffusive transport in the Fibonacci model at small $U$. Interactions at the mean-field level in fact \textit{always} slow down transport irrespective of $h$ value. This is manifestly untrue for the exact interacting dynamics where we have seen that 
interactions either ``enhance'' noninteracting subdiffusive transport to diffusion or ``degrade'' noninteracting superdiffusive transport to diffusion.
However in the case of binary disorder, there is no magnetization transfer in the absence of interactions due to Anderson localization; adding interactions at the mean-field level results in subdiffusive transport 
for the parameters chosen in Fig. \ref{fig:DWmf-bin}. Moreover we notice that the dynamical exponent in the presence of mean-field interactions is about the same for 
both types of disorder at a given $U, h$ combination (Fig.~\ref{fig:DWmf} and \ref{fig:DWmf-bin}). From this we conclude that a mean-field treatment is insufficient to discriminate the two types of disorder.

We note that in Ref.~\onlinecite{knap18} the method used (called a self-consistent Hartree-Fock method) is essentially the same as the mean field version (b), Eq.~(\ref{eq:mf2}). 
At stronger quasiperiodic potential and strong $U$ they find, differently than us at small $U$, that the above mean-field description leads to faster than subdiffusive relaxation. 
The applicability of the mean-field equations (\ref{eq:C},\ref{eq:mf1},\ref{eq:mf2}) needs to be studied in more detail.

\section{Conclusions}

Localization can be attributed to the physics of detuning between the strengths of off-diagonal to diagonal matrix elements in a local basis, whether in a random or quasiperiodic system \cite{Khemani17}.
Strong spatial correlations of the potential in the Fibonacci model prevent any detuning-caused localization in the non-interacting system at finite potential strength $h$. 
Upon introducing weak interactions we find no many-body localized phase and the transport of spin fluctuations becomes \textit{diffusive}, $\beta=\frac{1}{2}$, bearing no remnant of the spectral-fractality induced anomalous 
transport in the non-interacting limit. 
This latter aspect might be surprising as it suggests that already a small \textit{coherent} interaction is sufficient to wash out multifractality-induced anomalous diffusion in the noninteracting limit 
(noting that mean-field analysis suggests that multifractality should still persist in the Fibonacci model with weak interactions \cite{HiramotoMF}). Diffusive transport in the Fibonacci system must be contrasted with the binary disorder case, where there are no spatial correlations, and which shows instead a subdiffusive dynamics at the same field-strengths.

In this regard, there are two key differences here from the widely studied interacting-AAH model that make the interacting Fibonacci model interesting and worth pursuing further: 
Firstly, the resulting diffusion constant decays only algebraically with field strength as opposed to the Aubry-Andre-Harper model \cite{pnas18} where there is an exponential decay. Qualitatively, 
this is in line with the understanding that the cosine quasiperiodic potential is less correlated than the Fibonacci one, as reflected for instance in the existence of a non-interacting localization transition, 
and therefore shows slower transport, albeit being diffusive in both cases. 
However the precise reasons for this difference needs to be understood more quantitatively and is an interesting, open question.
Secondly, whereas the Fermi's golden rule estimate $\nu = 4(\beta-\frac{1}{2})$ for the diffusion constant scaling $D \sim U^{-\nu}$ for weak interactions $U$ 
gives a reasonable quantitative agreement with numerics, this estimate fails in the AAH model~\cite{pnas18}. On a similar note, mean-field analysis also suggests stronger discontinuity of spectral multifractality in the AAH model upon introducing weak interactions~\cite{HiramotoMF}.
These points suggest that the Fibonacci model is likely amenable to a more refined perturbative treatment for understanding its transport and (if any) localization properties.

Special care is dedicated to demonstrate a quantitative agreement between transport extracted from unitary dynamics of polarized domain walls or from boundary-driven Lindblad equation steady states, 
lending credence to our finding. 
We also show that mean-field analysis is unable to account for diffusion in the quasiperiodic model.

Finally, at larger interaction strength $U=0.5$ and $h=2$, we found a regime in parameter space where on available times $t \sim 800$ and system sizes $L \sim 144$ weak \textit{subdiffusive} dynamics occurs.
This finding, grossly agreeing with recent work \cite{Alet18}, is surprising: firstly, because the system is far away from the noninteracting limit such that multifractal effects of eigenfunctions and spectrum 
(which resulted in anomalous transport in the noninteracting limit \cite{Piechon}) should not play a role here, and secondly because there are no rare-regions which typically are invoked to account for 
subdiffusive transport \cite{agarwal15}. 
The full explanation of this effect requires further elucidation, starting with uncovering a possible similar effect in the interacting Aubry-Andre-Harper model by exploring a larger portion of 
its parameter space.

\section*{Acknowledgements}
VKV acknowledges support from the NSF DMR Grant No. 1508538 and US-Israel BSF Grant No. 2014265, and thanks N. Mace and V. Oganesyan for discussions.
MZ acknowledges Grants No.~J1-7279 and program No. P1-0402 of Slovenian Research Agency, and ERC AdG 694544 OMNES.

\appendix

\section{Finite size effects and log-periodic oscillations in the free system}

\begin{figure}
  \includegraphics[width=0.95\linewidth]{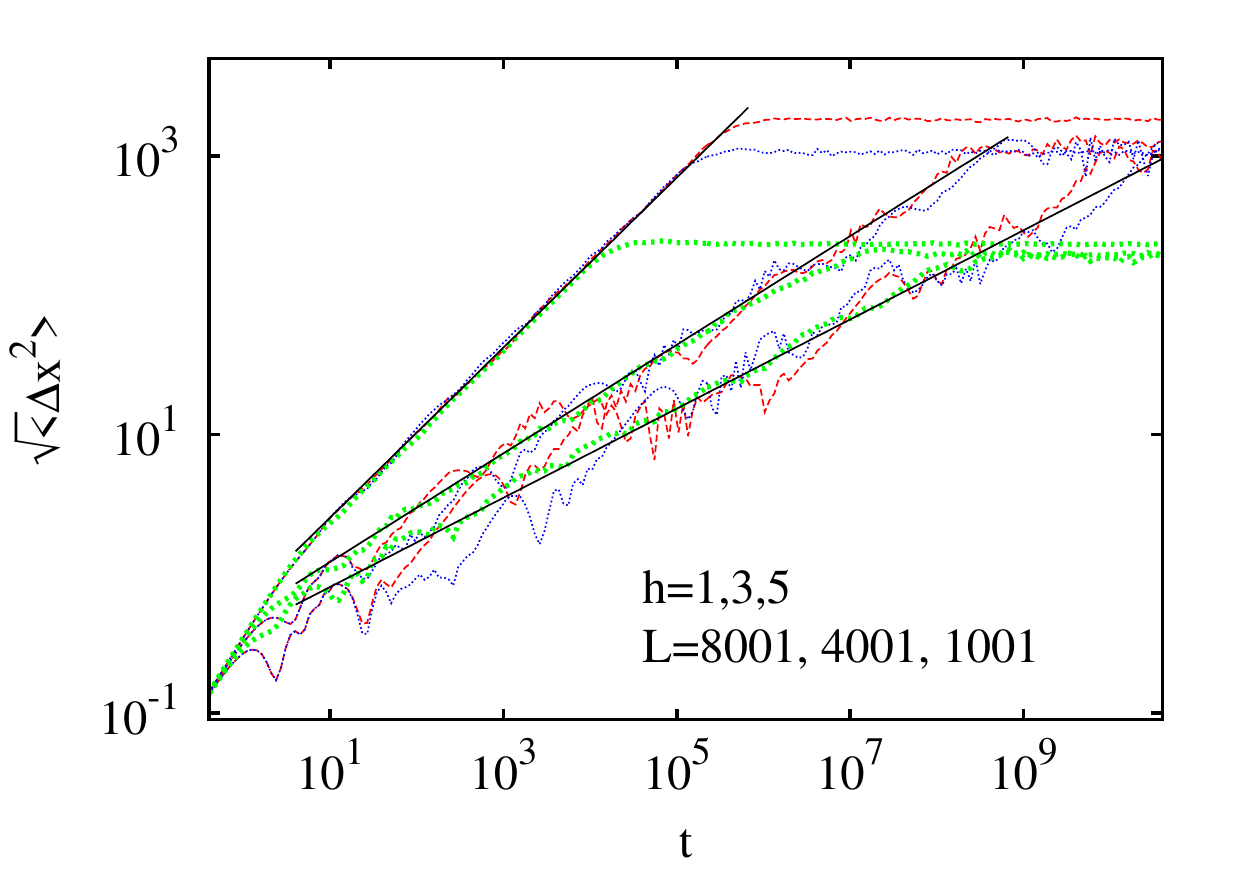}
  \hspace{2em}
  \includegraphics[width=0.92\linewidth]{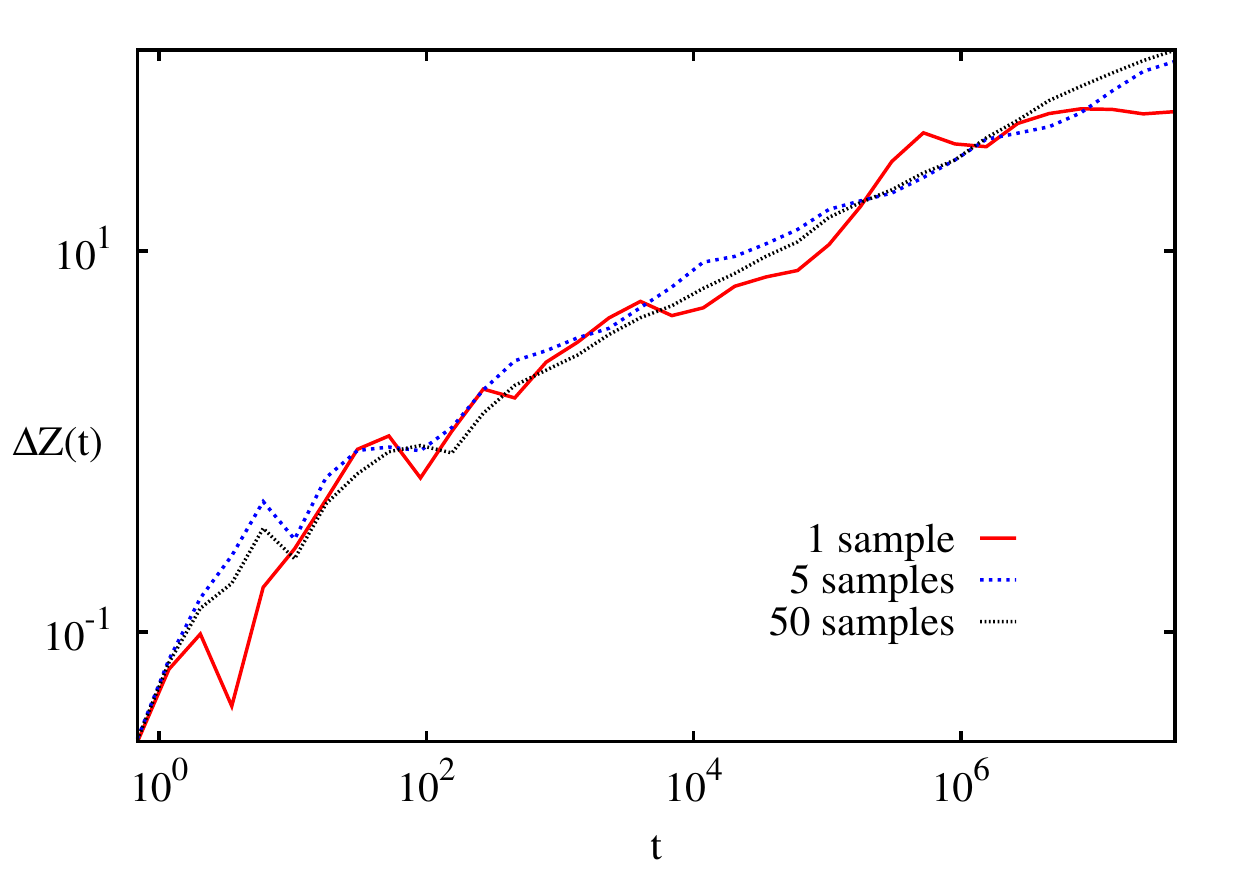}
 \caption{Top panel: Finite size effects in dynamics for $L=8001, 4001, 1001$ in the noninteracting Fibonacci system initialized with a delta function at the chain's centre.
 Bottom panel: Ensemble averaging effects, shown here for fixed $L=1000$ and $h=3$ for domain wall initial conditions with each curve averaged over different number of samples as indicated. 
 }
 \label{fig:finsize}
\end{figure}

In the main text we showed data from $L=8001$ for a single sample. Here we show that there are no finite size effects by going to smaller system sizes or ensemble averaging 
i.e. the average asymptotic growth may be reliably computed from a single sample in a large $L$ system.
In Fig. \ref{fig:finsize} we display the dynamics for a wavepacket spreading from a central site for $L=8001, 4001, 1001$. We see that they nicely follow each other (before finite-L saturation sets in), and 
that the dynamical exponent may be unambiguously determined.

\begin{figure}
  \includegraphics[width=1.\linewidth]{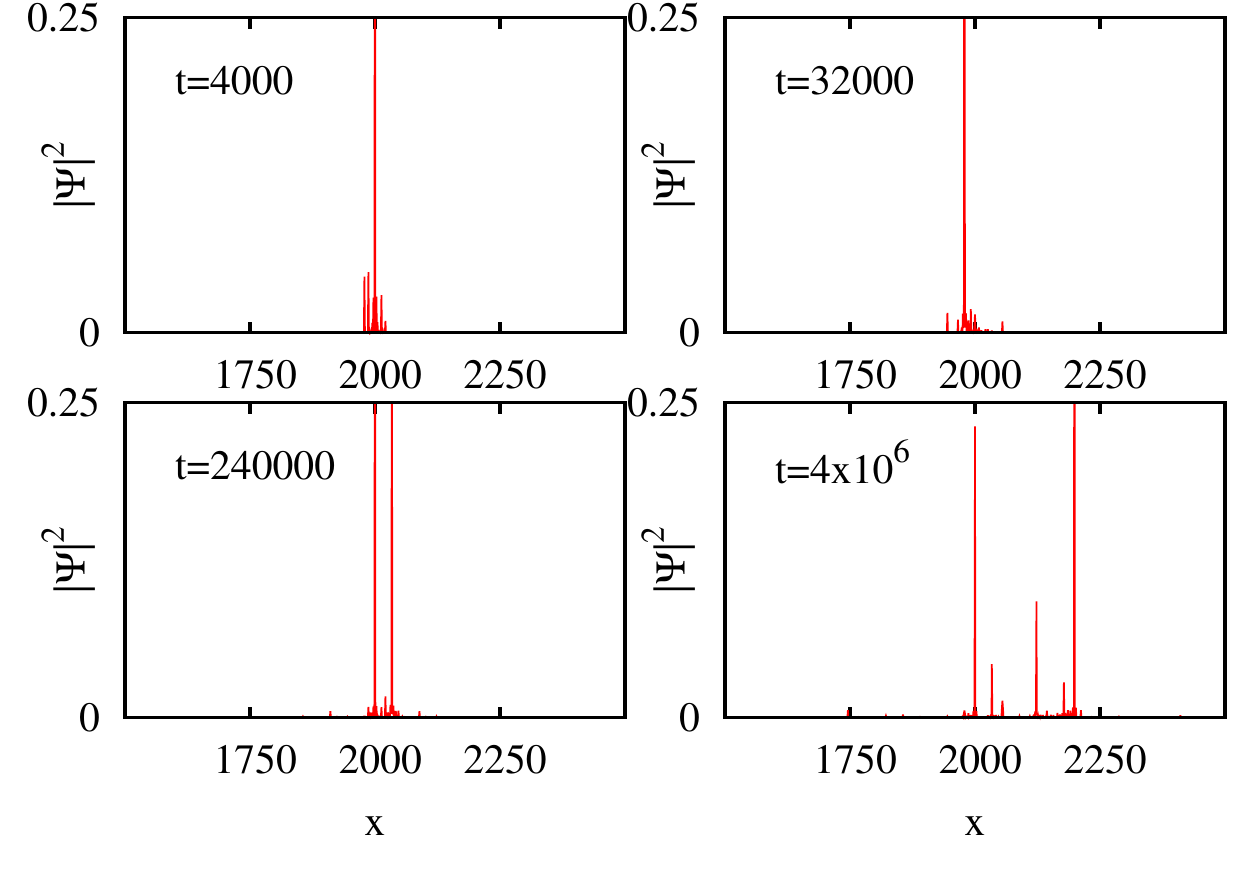}
   \hspace{2em}
   \includegraphics[width=1.\linewidth]{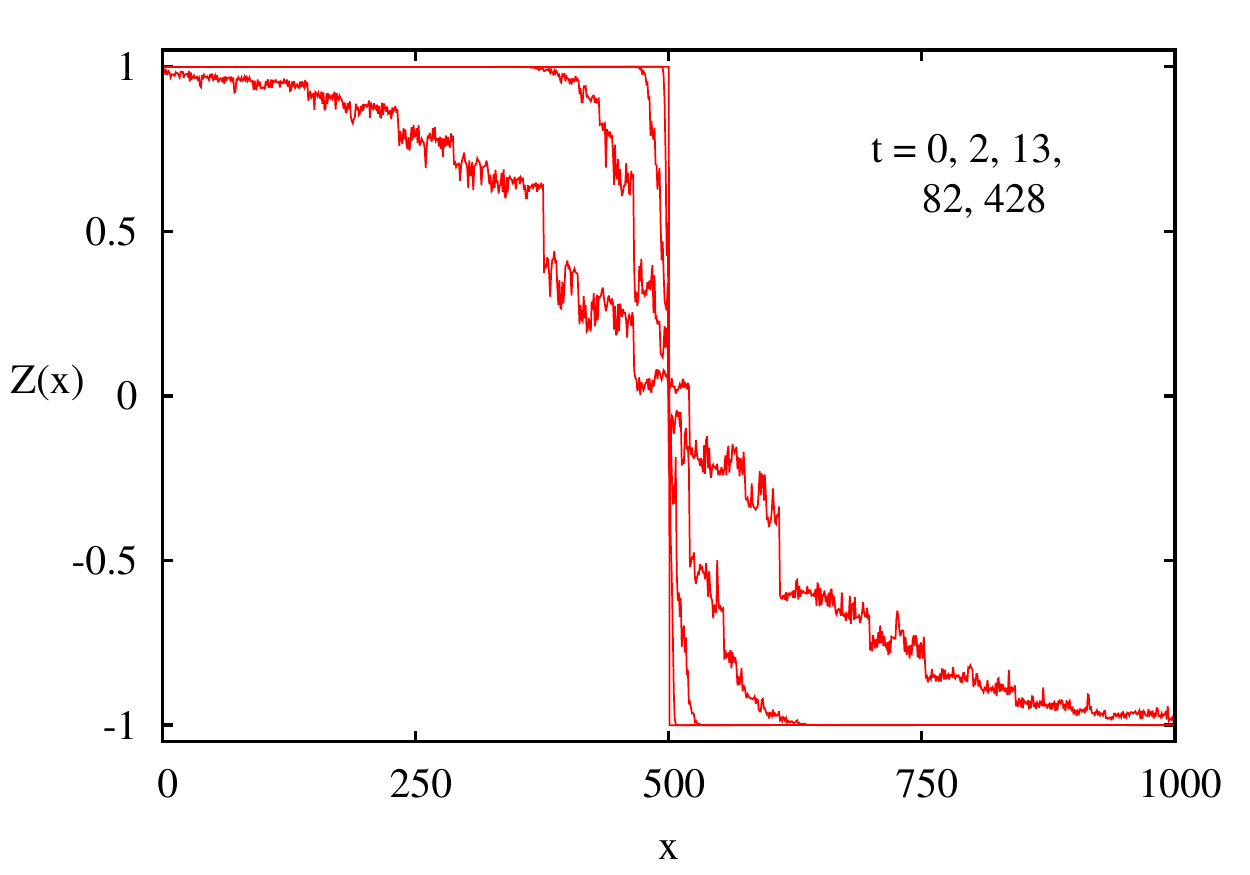}
 \caption{Top panel: The wavefunction amplitudes across the middle half of the chain at the indicated fixed times for $L=4001$ chain at $h=4$. 
 As time increases, we see that the wavefunction peak splits into two, developing a hierarchy; this self-similar hierarchy (entailing back and forth motion) is reflected in the above oscillations.
 Bottom panel: Magnetization profiles of domain-wall in the noninteracting $L=1000$ system at fixed times with $h=0.6$ and a single sample. 
 Here too we see a self-similar hierarchy developing in the profiles.}
 \label{fig:nonintprofiles}
\end{figure}

However around the average algebraic growth (whether for wavepacket spreading or transferred magnetization, as we see in Fig. \ref{fig:finsize} and Fig. \ref{fig: free}), there are oscillations around this 
anomalous dynamics especially at larger $h$ values. These oscillations result from the self-similar hierarchy developing in the profiles (see Fig. \ref{fig:nonintprofiles}) which in turn arise from the spectral fractality 
and the subsequent oscillatory transferring of wavefunction weight between different peaks for the wavepacket spreading \cite{abe1987, wilkinson1994, zhong1995, yuan2000, thiem2009}.

For the domain wall spreading, we find a similar story; see bottom panel of Fig. \ref{fig:nonintprofiles}. There we see a hierarchical development of mini domain-walls along the chain, causing 
recurring magnetization fluctuations along these new domain walls, resulting in oscillations in the dynamics. 
However due to the bulk nature of the initial excitations (requiring that excess magnetization be transferred across half-chain lengths in order for a true reversal or oscillation in the dynamics), 
these oscillations are significantly weaker here than in the wavepacket spreading.

\section{Further NESS data and the convergence with the MPO bond dimension}
\label{app:convergence}

\begin{figure}[ht!]
\centerline{\includegraphics[width=0.9\linewidth]{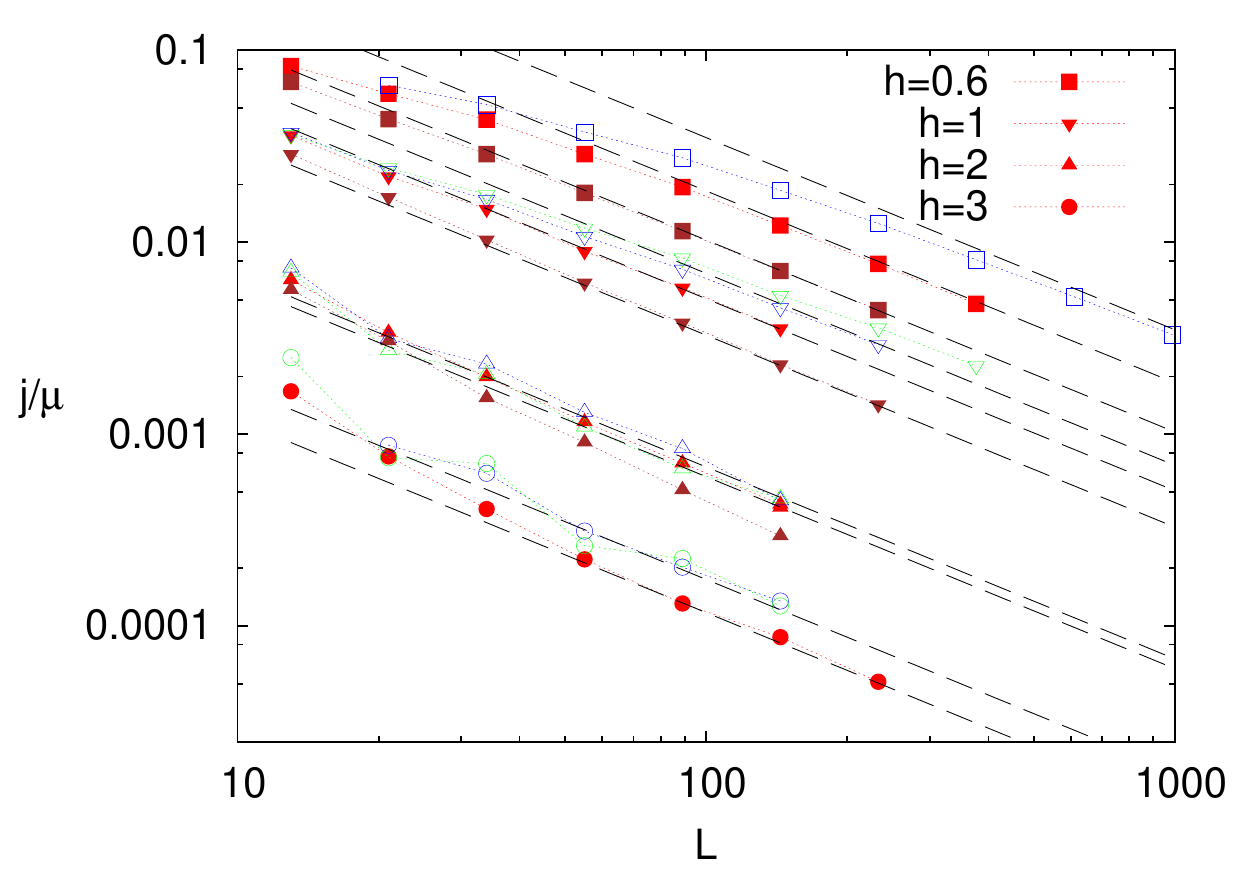}}
\caption{Scaling of the NESS magnetization current with system size $L$. Dashed lines are diffusive $D_{\rm NESS}/L$ (see Table~\ref{tab1}), squares, down triangles, triangles and circles are for $h=0.6,1,2,3$, respectively. Green color is for $U=0.05$, blue for $U=0.1$, red for $U=0.2$, and brown for $U=0.5$. In all cases one has asymptotic diffusion, except for $h=2$ and $U=0.5$ (brown triangles). All data is for one Fibonacci realization.}
\label{fig:tok}
\end{figure}

In Fig.~\ref{fig:tok} we show data for the NESS current, from which by fitting $1/L$ dependence we extract the NESS diffusion constant $\Dn$ (Table~\ref{tab1}).

\begin{figure}[ht!]
\centerline{\includegraphics[width=\linewidth]{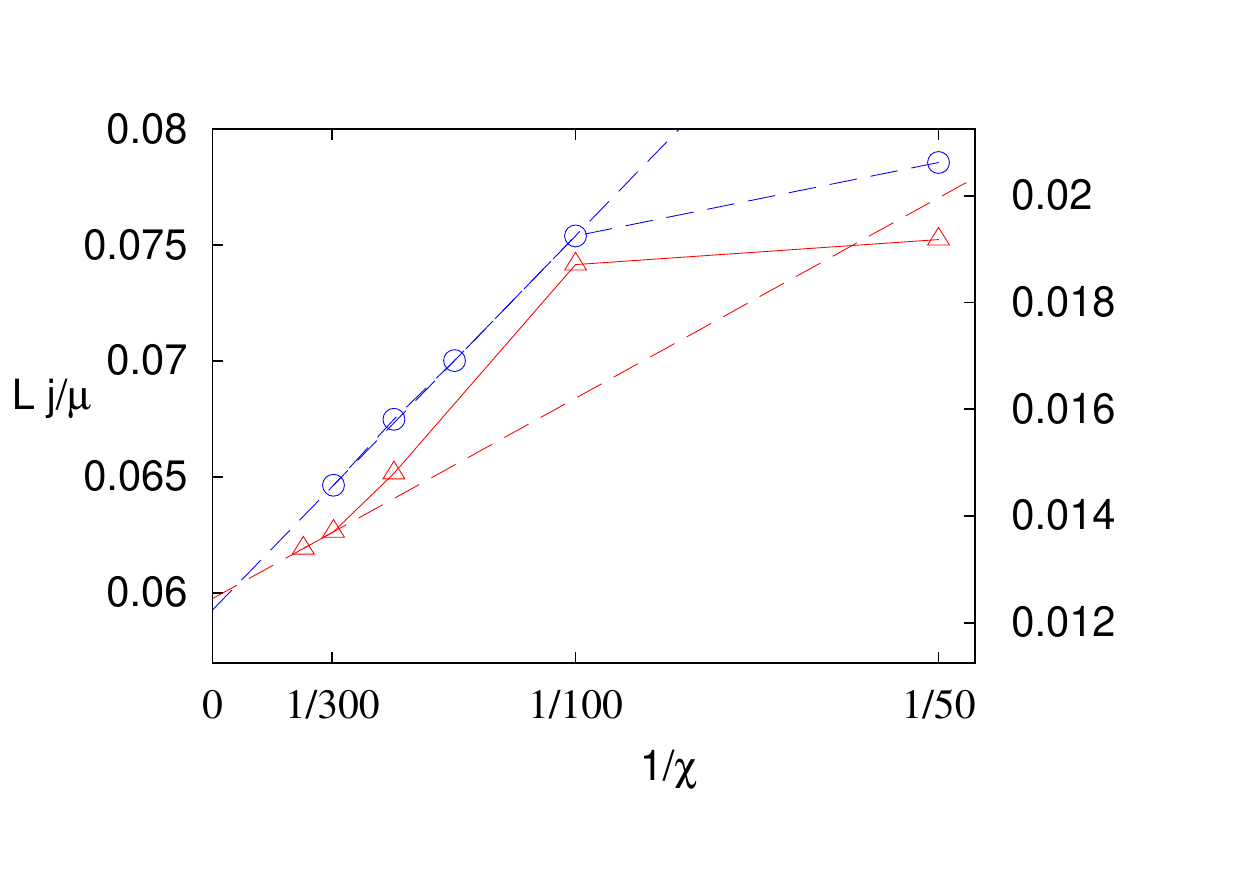}}
\caption{Convergence of NESS current $j$ with MPO bond dimension $\chi$. We show one (typical) case of nice convergence for $U=0.2$, $h=3$ and $L=55$ (blue color, right axis), and one case of less nice convergence obtained for $U=0.2$, $h=2$ and $L=144$ (red color, left axis). Dashed lines show the used extrapolation to $\chi \to \infty$.}
\label{fig:extrapol}
\end{figure}

As mentioned, we always check convergence with the MPO bond dimension $\chi$. Typically the NESS current $j$ decreases as one increases $\chi$, with a finite-$\chi$ correction scaling in most cases as $1/\chi$. In cases where the required $\chi$ in order to get precision of order (few) percent would be prohibitively large we therefore use extrapolation in $1/\chi$ in order to get closer to a true $j$. This is shown in Fig.~\ref{fig:extrapol}.
One can see that a finite $\chi$ introduces an additional ``truncation noise'' that makes transport faster, i.e., increases $j$.

\begin{figure}[ht!]
\centerline{\includegraphics[width=\linewidth]{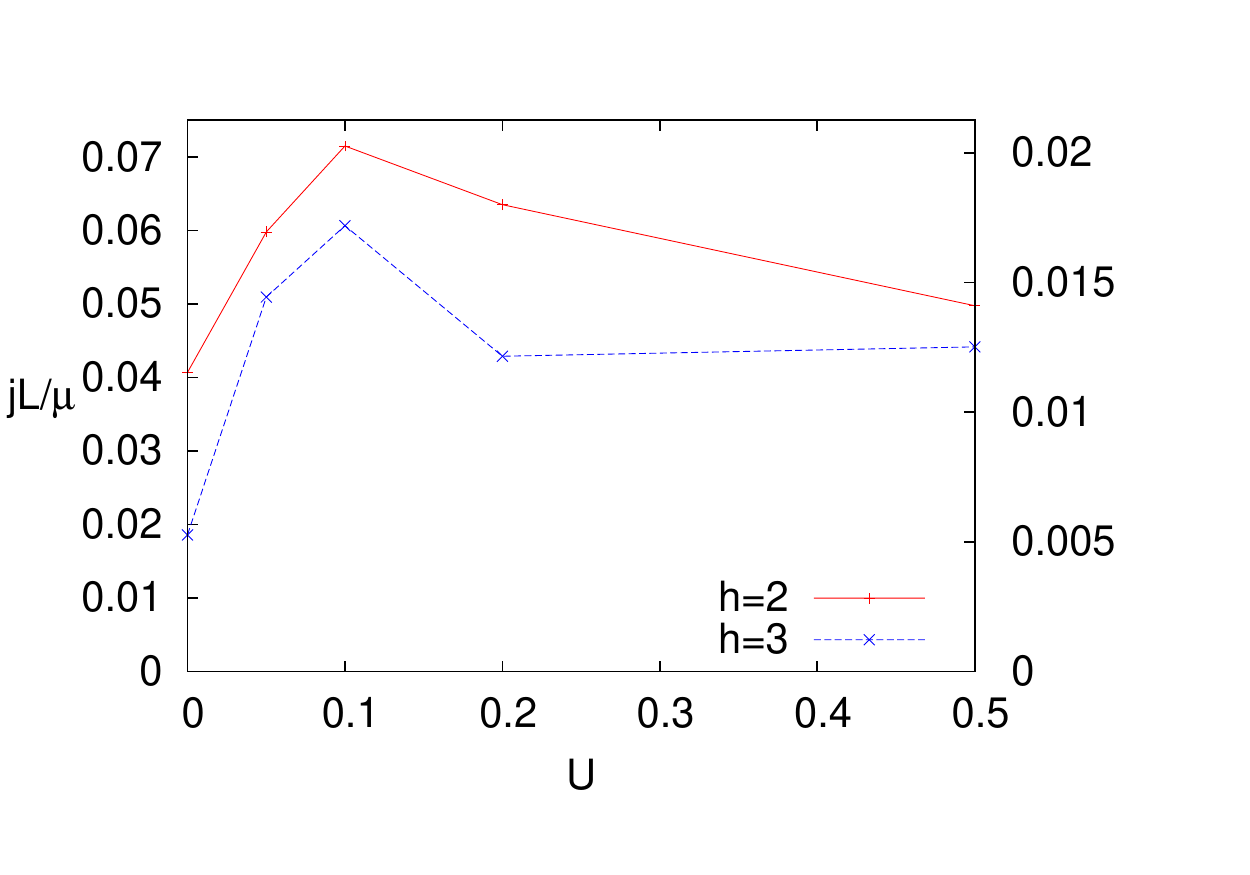}}
\caption{Scaled NESS current for $L=55$ and $h=2$ (red, left axis), and $h=3$ (blue, right axis). For both values of $h$ one has subdiffusion at $U=0$. As a function of interaction $U$ one has a non-monotonous dependence.}
\label{fig:maxU}
\end{figure}

In Fig.~\ref{fig:maxU} we show the dependence of $j$ on interaction for a fixed $L$ and $h\ge 2$ where the noninteracting system is subdiffusive.
We can see that the current initially, expectedly, increases with $U$, reaching a maximum around $U \approx 0.1$. Such small value of the maximum prevents us to explore more in detail a conjectured power-law dependence at small $U$, that is to extract the scaling exponent $\nu$ for $h\ge 2$.

\section{Additional unitary data}

\begin{figure}[ht!]
\centerline{\includegraphics[width=0.9\linewidth]{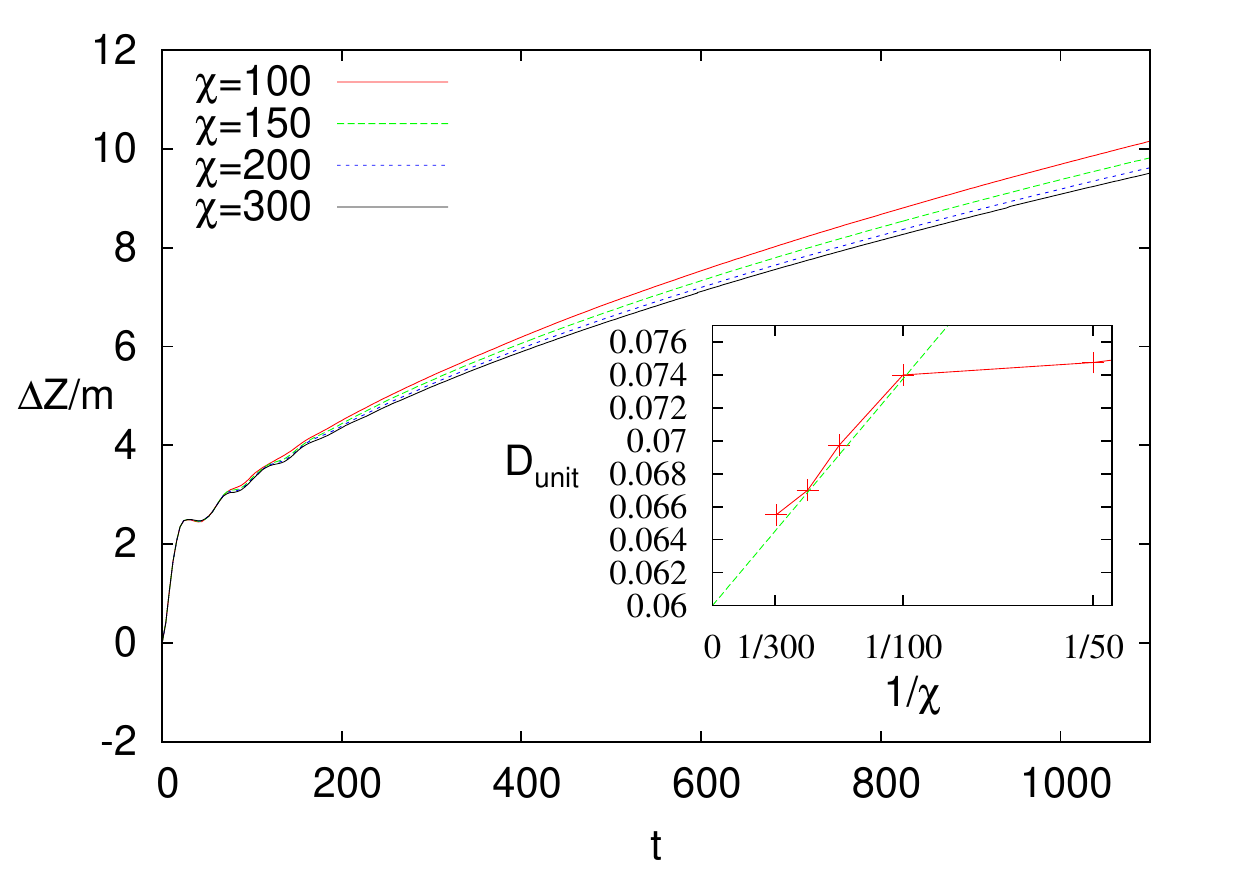}}
\caption{Convergence with $\chi$ for unitary evolution of a weakly polarized domain wall. The inset shows the convergence of $\Du$ (obtained by fitting $(4\Du t/\pi)^{1/2}$ to each curve for $t \in [400,1200]$). All is for $U=0.2$, $h=2$, and $L=144$ (one Fibonacci sequence).}
\label{fig:convunit}
\end{figure}

In Fig.~\ref{fig:convunit} we show data for different $\chi$ and unitary evolution of a weakly polarized domain wall (\ref{eq:dw}). We can see that the convergence of $\Du$ is similarly $\sim 1/\chi$ as for the NESS setting.

\begin{figure}[ht!]
\centerline{\includegraphics[width=0.9\linewidth]{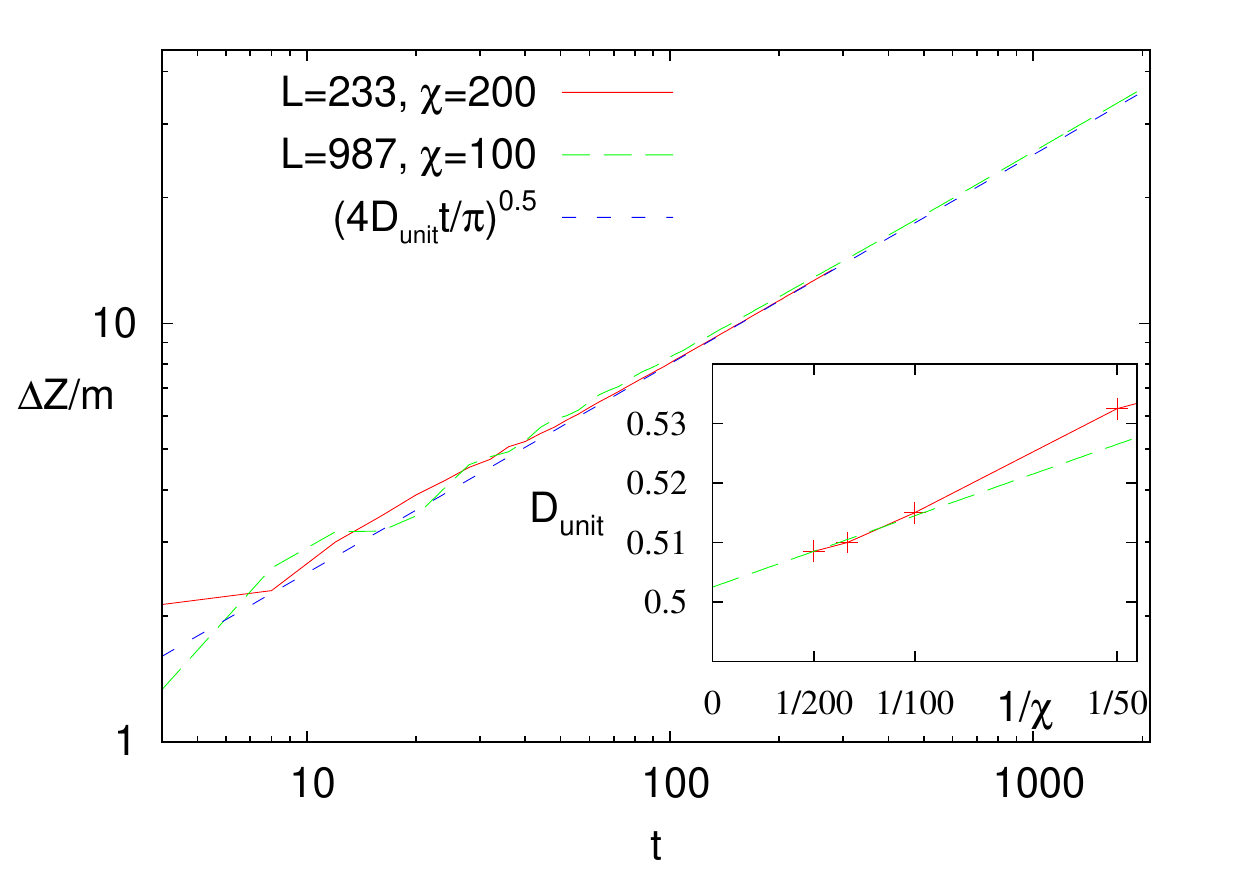}}
\caption{Unitary evolution of a weakly polarized domain wall for different Fibonacci sequences (i.e., different $L$). One can see that asymptotically $\Du \approx 0.50$ is independent of $L$ and potential realization. All is for $U=0.2$, $h=1$ (one Fibonacci sequence).}
\label{fig:comp}
\end{figure}

In Fig.~\ref{fig:comp} we show data for different $L$ and therefore different Fibonacci sequence realizations. While at short times there is understandably a difference between realizations (red and green data in Fig.~\ref{fig:comp}), at longer times diffusion becomes ``self-averaging'' with very little difference between different Fibonacci sequence realizations. This is also a reason that we typically do not have to do any averaging. Comparing $h=1$ (Fig.~\ref{fig:comp}) and $h=2$ (Fig.~\ref{fig:convunit}), the absolute size of finite-$\chi$ correction is about the same (about $\approx 0.03$ at $\chi=200$), however, the relative error is for $h=1$ almost $10$ times smaller than for $h=2$ simply because of larger $\Du$. This means that smaller $\chi$ suffices at $h=1$ for the same relative precision of $\Du$. This is the reason we could afford to go to $t=2000$ and $L=987$ for smaller $h=1$, and why on the other hand simulations for larger $h$ are more time consuming.


\begin{thebibliography}{100}
\bibitem{foot3} They however mostly do not agree on the values of exponents, with some of that discrepancy probably being due to small systems/times used in many studies.
\bibitem{foot00} We could afford to use $\chi \le 300$ (in few cases, e.g. NESS for $U=0.2$, $h=2$, even $\chi=400$, or $\chi=500$ for unitary $U=0.5$, $h=2$). Beware that a mixed state evolution using tDMRG is about $8$ times costlier than a pure state evolution (each short time-step involves a SVD of a $4\chi \times 4\chi$ matrix instead of a $2\chi \times 2\chi$). The total CPU computational time for all the data gathered was more than $10^6$ core-hours!
\bibitem{foot0} It makes sense to speak about tranport type only in the limit of $t \to \infty$, or $L \to \infty$ for the NESS setting. Only in such limit can one unambgiously distinguish different scalings $\sim t^\alpha$.
\bibitem{foot1} The claim made in Ref.~\onlinecite{barlev17} about superdiffusion for small $h$ as well as for small $U$, based on short and narrow range of times (note that the relevant parameter is the number of scattering events $U^2t$, so a priori taking just smaller $U$ does not help), are both incorrect, as well as the statement that there is no finite region of parameters with diffusion. In fact, it is the opposite, one always has diffusion, see Ref.~\onlinecite{pnas18} for correct statements.
\end{thebibliography}
\end{document}